%% file: main.tex
\documentclass[journal]{IEEEtran}
\usepackage{cite}
\usepackage{graphicx, subfig, caption}
\graphicspath{{../figures/}}
\usepackage{amsmath}
\usepackage{amssymb}
\usepackage{amsthm}
\newtheorem{remark}{Remark}
\usepackage{bbm}
\usepackage{bm}
\usepackage[T1]{fontenc}

\usepackage[acronym,nogroupskip,nonumberlist,nopostdot]{glossaries}
\usepackage{xspace}
\usepackage{color}
\usepackage{algorithm}
\usepackage[noend]{algorithmic}
\usepackage{hyperref}

\glsdisablehyper
\input{glossary}

\input{notations}

\allowdisplaybreaks
\begin{document}
\bstctlcite{IEEEexample:BSTcontrol}
\title{Harnessing Multimodal Sensing for Multi-user Beamforming in mmWave Systems}
\author{Kartik~Patel,~\IEEEmembership{Graduate Student Member,~IEEE,}
        Robert W.~Heath Jr.,~\IEEEmembership{Fellow,~IEEE}
\thanks{
Kartik Patel is with the Chandra Department
of Electrical and Computer Engineering, the University of Texas at Austin, Austin, TX, 78712 USA. Email: \texttt{kartikpatel@utexas.edu}.
Robert W. Heath Jr. is with the department of Electrical and Computer Engineering, University of California San Diego, San Diego, CA. Email: \texttt{rwheathjr@ucsd.edu}.
This material is based upon work supported by the National Science Foundation under grant nos. NSF-ECCS-2153698, NSF-CCF-2225555, NSF-CNS-2147955 and is supported in part by funds from federal agency and industry partners as specified in the Resilient \& Intelligent NextG Systems (RINGS) program.
}
}

\maketitle

\begin{abstract}
Sensor-aided beamforming reduces the overheads associated with beam training in \mmWave \MIMO communication systems. Most prior work, though, neglects the challenges associated with establishing \MU communication links in \mmWave \MIMO systems. In this paper, we propose a new framework for sensor-aided beam training in \MU \mmWave \MIMO system. We leverage the beamspace representation of the channel that contains only the \AoDs of the channel's significant multipath components. 
We show that a \DNN-based multimodal sensor fusion framework can estimate the beamspace representation of the channel using sensor data. To aid the \DNN training, we introduce a novel \SSCL function that leverages the inherent similarity between channels to extract similar features from the sensor data for similar channels. 
Finally, we design an \MU beamforming strategy that uses the estimated beamspaces of the channels to select analog precoders for all users in a way that prevents transmission to multiple users over the same directions. Compared to the baseline, our approach achieves more than 4$\times$ improvement in the median sum-\SE at 42 dBm \EIRP with 4 active users. This demonstrates that sensor data can provide more channel information than previously explored, with significant implications for \ML-based communication and sensing systems.
\end{abstract}

\section{Introduction}

Hybrid \MIMO architectures enable large antenna arrays to support high-bandwidth applications like \mmWave communication. Configuring communication links with hybrid \MIMO architectures in commercial systems like 5G and IEEE 802.11ay make use of beam training protocols~\cite{DahParSko:5G-NR:-The-Next-Generation-Wireless:20,GioPolRoy:A-Tutorial-on-Beam-Management:19,DreHea:Massive-MIMO-in-5G:-How-Beamforming:23,Ghada-Cor:IEEE-802.11ay:-Next-Generation:17}. These protocols involve a multi-stage procedure where both the transmitter and the receiver measure candidate beams from codebooks. The measurements are used to eventually configure the analog (\RF) and digital precoders and combiners for multi-stream \MIMO transmission. While precoder and combiner search occur in a multi-stage hierarchical fashion, it remains a source of high overhead in configuring MIMO communication links.

Prior work has proposed using out-of-band sensor data to reduce the beam training overhead by narrowing the set of candidate beam pairs to try. These sensor data include position information~\cite{AviKou:Position-aided-mm-wave-beam:16,GonAliVa:Millimeter-Wave-Communication-with:17,LocAsaSim:mmWave-on-wheels:-Practical:17,MorBehPez:Position-Aided-Beam:22,VaChoShi:Inverse-multipath-fingerprinting:17}, radar signals~\cite{AliGonGho:Passive-Radar-at-the-Roadside:20,GonRiaHea:Radar-Aided-Beam:16,AliGonHea:Leveraging-Sensing-at-the-Infrastructure:20,DemAlk:Radar-Aided-6G-Beam:22,GraCheGon:Deep-Learning-Based-Link:23,ZhaPatSha:Side-information-aided-Noncoherent-Beam:19}, camera images~\cite{AlrBooHre:ViWi-Vision-Aided-mmWave:20,AlrHreAlk:Millimeter-Wave-Base:20,SalBelSan:Machine-Learning-on-Camera:20,TiaPanAlo:Applying-deep-learning-based-computer:20} and LiDAR point clouds~\cite{JiaChaAlk:LiDAR-Aided-Future:22,KlaGonHea:LIDAR-Data-for-Deep:19,WooZhaCha:SpaceBeam:-LiDAR-driven-one-shot:21}. While some early work used mathematical techniques~\cite{GonRiaHea:Radar-Aided-Beam:16,AliGonGho:Passive-Radar-at-the-Roadside:20}, most recent approaches leverage \ML to learn the relationship between sensor information and the optimal beam configuration. Further fueled by the publicly available datasets like Raymobtime~\cite{KlaBatGon:5G-MIMO-Data-for-Machine:18}, e-FLASH~\cite{GuSalRoy:Multimodality-in-mmWave-MIMO:22} and DeepSense 6G~\cite{AlkChaOsm:DeepSense-6G:-A-Large-Scale-Real-World:23}, prior work has also explored using a combination of position information, camera images, and LiDAR point clouds to improve the accuracy of sensing-aided beam prediction strategies~\cite{DiaKlaGon:Position-and-LIDAR-Aided-mmWave:19,SalReuRoy:Deep-Learning-on-Multimodal:22,ChaOsmHre:Vision-position-multi-modal-beam:22,RoySalBan:Going-beyond-RF:-A-survey:23,ChaHreSto:Towards-real-world-6G-drone:22, ReuSalRoy:Deep-Learning-on-Visual:21}. 

\begin{figure*}
    \centering
    \subfloat[\label{fig:overview-a}]{\includegraphics[width = .4\textwidth]{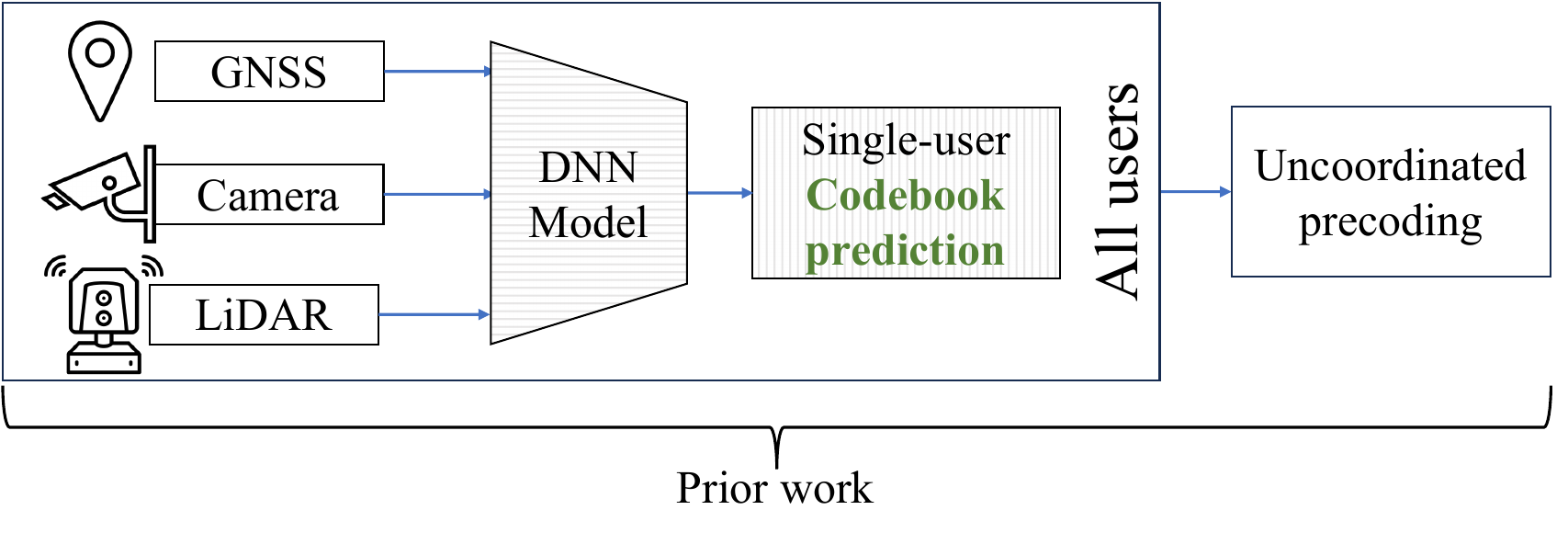}}\quad\quad\quad
    \subfloat[\label{fig:overview-b}]{\includegraphics[width = .4\textwidth]{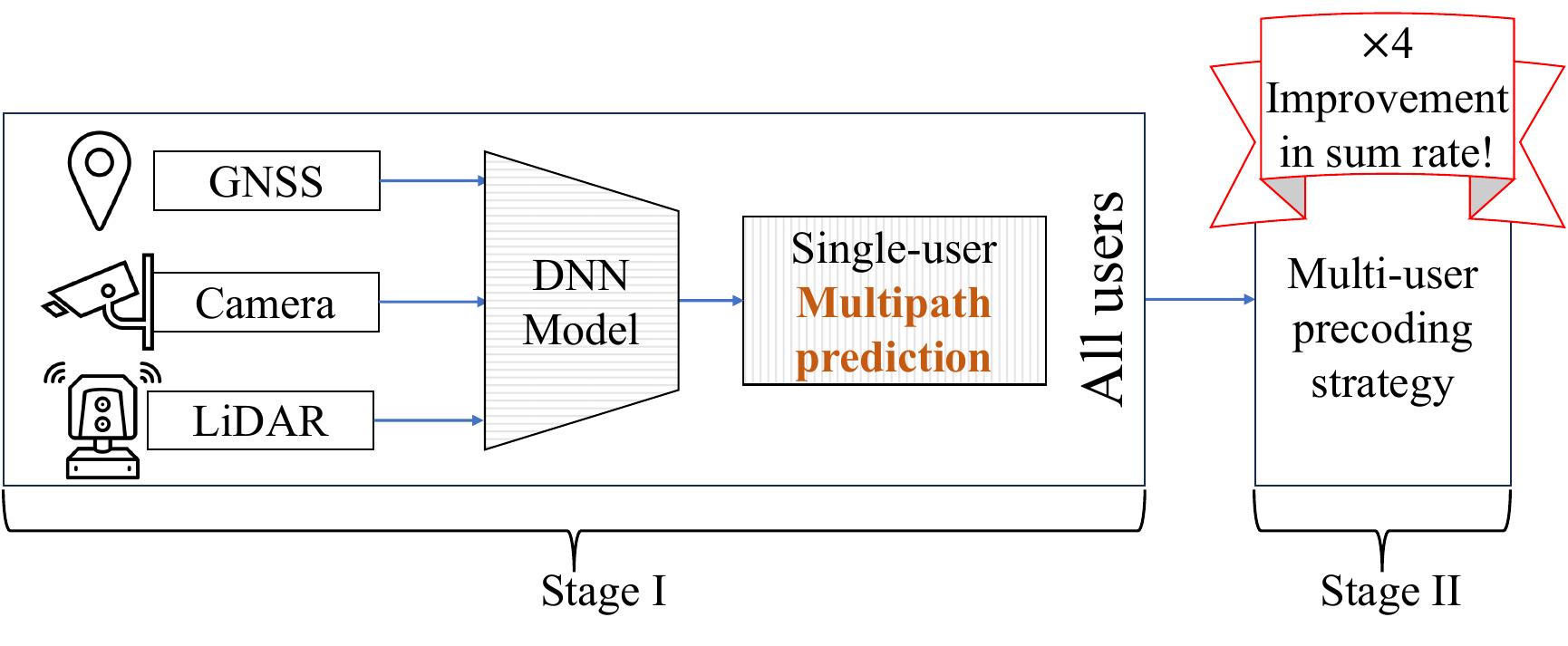}}
    \caption{Improving \MU communication using side-information: (a) Prior work uses single-user codebook prediction that could suffer from inter-user interference in \MU setting. We show that predicting multipath components and using a dedicated beamforming strategy designed for \MU scenario can double the achievable rates for \MU systems.}
    \label{fig:overview}
\end{figure*}

While existing \ML-based approaches~\cite{DiaKlaGon:Position-and-LIDAR-Aided-mmWave:19,SalReuRoy:Deep-Learning-on-Multimodal:22,ChaOsmHre:Vision-position-multi-modal-beam:22,RoySalBan:Going-beyond-RF:-A-survey:23,ChaHreSto:Towards-real-world-6G-drone:22, ReuSalRoy:Deep-Learning-on-Visual:21,AviKou:Position-aided-mm-wave-beam:16,GonAliVa:Millimeter-Wave-Communication-with:17,LocAsaSim:mmWave-on-wheels:-Practical:17,MorBehPez:Position-Aided-Beam:22,VaChoShi:Inverse-multipath-fingerprinting:17,AliGonHea:Leveraging-Sensing-at-the-Infrastructure:20,DemAlk:Radar-Aided-6G-Beam:22,ZhaPatSha:Side-information-aided-Noncoherent-Beam:19,AlrBooHre:ViWi-Vision-Aided-mmWave:20,AlrHreAlk:Millimeter-Wave-Base:20,SalBelSan:Machine-Learning-on-Camera:20,TiaPanAlo:Applying-deep-learning-based-computer:20,JiaChaAlk:LiDAR-Aided-Future:22,KlaGonHea:LIDAR-Data-for-Deep:19,WooZhaCha:SpaceBeam:-LiDAR-driven-one-shot:21} show promise in the \SU setting, their performance in the \MU setting remains unclear (as summarized in \figref{fig:overview-a}). Specifically, predicting the optimal beams for each \UE separately might overlook inter-user interference; a crucial aspect of \MU communication. Therefore, a framework is needed that is specifically designed for \MU setting that leverages sensor data for low-overhead beam training while considering inter-user interference. 

Some prior work has explored \ML-based solutions for \MU beam training using the history of \CSI and signal strength measurements~\cite{DreHea:Machine-Learning-Codebook:23, MyeWanGon:Deep-Learning-Based-Beam:20, MaQiLi:Machine-Learning-for-Beam:20, MolArd:Deep-Learning-Assisted:22, HuaYeXia:Hybrid-Beamforming-for-Millimeter:20}. There is, however, very limited work that leverages sensor data for \MU beam training. One approach uses camera images for \MU beam training and proposes a two-stage process with user selection and beam training stages, each using separate \DNN models~\cite{AhnOriKan:Machine-Learning-Based-Vision-Aided:22}. That approach, however, is limited by an assumption of the existence of a single dominant \LOS path to each user, which is not realistic in practice~\cite{JaiKumPan:The-Impact-of-Mobile-Blockers:19}. 

In this paper, we propose using multimodal sensor data, including camera images, LiDAR point clouds, and position information, for beam training in \MU \mmWave systems (see \figref{fig:overview-b}). We hypothesize that this combined sensor data offers more detailed information about \MU-\MISO \mmWave channels than just dominant paths and blockages, as assumed in prior work~\cite{AviKou:Position-aided-mm-wave-beam:16,GonAliVa:Millimeter-Wave-Communication-with:17,LocAsaSim:mmWave-on-wheels:-Practical:17,MorBehPez:Position-Aided-Beam:22,VaChoShi:Inverse-multipath-fingerprinting:17,AliGonHea:Leveraging-Sensing-at-the-Infrastructure:20,DemAlk:Radar-Aided-6G-Beam:22,GraCheGon:Deep-Learning-Based-Link:23,ZhaPatSha:Side-information-aided-Noncoherent-Beam:19,AlrBooHre:ViWi-Vision-Aided-mmWave:20,AlrHreAlk:Millimeter-Wave-Base:20,SalBelSan:Machine-Learning-on-Camera:20,TiaPanAlo:Applying-deep-learning-based-computer:20,JiaChaAlk:LiDAR-Aided-Future:22,KlaGonHea:LIDAR-Data-for-Deep:19,WooZhaCha:SpaceBeam:-LiDAR-driven-one-shot:21,DiaKlaGon:Position-and-LIDAR-Aided-mmWave:19,SalReuRoy:Deep-Learning-on-Multimodal:22,ChaOsmHre:Vision-position-multi-modal-beam:22,RoySalBan:Going-beyond-RF:-A-survey:23,ChaHreSto:Towards-real-world-6G-drone:22, ReuSalRoy:Deep-Learning-on-Visual:21}. We show that these sensors can collectively provide information about the \AoDs of the channel's significant multipath components. 

To estimate the \AoDs of the channel paths from sensor data, we employ a \DNN-based multimodal fusion network similar to \cite{SalReuRoy:Deep-Learning-on-Multimodal:22}. Instead of identifying a single optimal beam, however, we aim to identify \AoDs of the channel paths. This requires transforming the problem from a classification to a multi-label identification problem. Thus, each quantized AoD pair (azimuth and elevation) is treated as a label. This approach enables a fixed size \DNN to predict the \AoDs of a variable number of multipath components. 

While the multi-label identification problem offers a basic methodology for \DNN training, it has a limitation. Conventional multi-label identification tasks assume categorical labels with no notion of \textit{proximity} between labels. The labels can only be the same or different. Therefore, the samples are similar only if they share a few identical labels in the ground truth. In contrast, \AoDs of channel paths have an inherent notion of proximity -- they can be closer or farther apart, not just the same or different. Consequently, there is also a notion of similarity between the channels, even when their channel paths do not share identical \AoDs. Furthermore, this notion of similarity between the channels can also enhance \DNN training by encouraging it to extract similar features from sensor data when the channels have similar \AoDs of their paths. 

In this paper, we formalize this key insight. Specifically, we quantify the notion of similarity using a novel \textit{soft encoding} of \AoDs of channel paths and, subsequently, defining a metric of similarity between two channels. We further introduce a novel loss function called \SSCL to aid \DNN training by incentivizing the intermediate features extracted by the \DNN network to align or dis-align in proportion to the similarity of the channels. \SSCL is inspired from \SCL~\cite{KhoTetWan:Supervised-contrastive-learning:20} which incentivizes the \DNN in a similar manner but is only designed for categorical classes without capturing the similarity across them. 
We believe \SSCL loss function can be valuable for applying supervised \ML algorithms to practical systems like wireless communication, where labels deal with non-categorical data and there exists a notion of similarity across labels. To the best of our knowledge, this is the first time such an encoding and loss function have been used for training supervised \ML models. 

Finally, we define the beamspace representation of the channel containing only \AoDs of the channel paths and propose a novel \MU beamforming algorithm based on sensor-aided beamspace estimates. This algorithm aims to mitigate interference by preventing transmissions to multiple users over the same multipath \textit{clusters}. We show that this strategy, specifically designed for \MU settings, achieves 4$\times$ improvement in the median sum-\SE with 4 active users compared to \MU extensions of \SU beam prediction strategies, and reduces the link establishment overhead. 

The major contributions of this paper are summarized as follows:
\begin{enumerate}
    \item We propose a framework for sensor-aided \MU beam training at the \BS. We use the beamspace representation of a channel, which focuses only on the \AoDs of channel paths. We design an \ML-based multimodal fusion network trained using the camera, LiDAR and position information, and the beamspaces of the associated channels. 
    By formulating the problem as a multi-label identification task with each label representing a quantized \AoD pair (azimuth and elevation), we enable the prediction of a variable number of \AoD pairs. 
    \item We propose a novel soft-encoding technique and a corresponding similarity metric that captures the similarity between channels based on the proximity of the \AoDs of their channel paths. Furthermore, we introduce the \SSCL, a novel loss function that incentivizes the network to extract similar features from the sensors for similar beamspaces. We believe \SSCL has broader applicability to various \ML tasks in wireless systems that often involve non-categorical labels. 
    \item We present a novel \MU beamforming algorithm that uses the estimated beamspace representations of the channels to identify the \RF precoders at the \BS. Subsequently, conventional preamble-based channel estimation techniques and \RZF are used to design the digital precoders. 
    \item We conduct a comprehensive analysis on the Raymobtime dataset~\cite{KlaBatGon:5G-MIMO-Data-for-Machine:18}, assessing the end-to-end performance of the proposed sensor-aided \MU beamforming against two baselines, \SU approaches extended to \MU setting and Full \CSI setting. This evaluation showcases the ability of sensor data to extract significantly more information about channels and the benefit of using a joint MU beamforming strategy.
\end{enumerate}
\noindent \textbf{Organization:} The rest of the paper is structured as follows: \secref{sec:system} contains the system model and formally defines the beamspace of a channel. In \secref{sec:stageI}, we detail the sensor-aided beamspace prediction process, including sensor data preprocessing, ground truth beamspace encoding, the \DNN architecture, and loss functions and metrics used for training and testing. In \secref{sec:stageII}, we describe the proposed \MU beamforming algorithm that leverages the estimated beamspaces of the channels and selects \RF and digital precoders to minimize inter-user interference across users. Finally, we present the performance analysis of the proposed sensor-aided \MU beamforming strategy in \secref{sec:results} and conclude in \secref{sec:conclusion}. 

\noindent \textbf{Notations}: Small bold letters ($\mathbf{a}$) and capital bold letters ($\mathbf{A}$) denote a vector and matrix, respectively. $\mathbf{\conj{a}}, \mathbf{a}^\trans, \mathbf{a}^*$ denotes conjugate, transpose, and Hermitian of the vector/matrix $\mathbf{a}$. 
$[a]$ denotes the set $\{0,\ldots,a-1\}$. 
$\vectorize{\mathbf{A}}$ denotes the vectorization operator on the matrix $\mathbf{A}_{M\times N}$ such that the $i,j$-th element of the matrix $\mathbf{A}$ maps to $(i N + j)$-th element of the column vector $\vectorize{\mathbf{A}}$, $\forall i\in[M], j\in[N]$. 
The hat $\hat{\mathbf{a}}$ denotes estimated value of $\mathbf{a}$. Bold $\mathbf{1}_{N}$ denotes the vector of length $N$ with all elements equal to one and capital bold $\mathbf{I}_{N}$ denotes the identity matrix of dimensions $N\times N$. Finally, $\angdiff{\theta_1}{\theta_2}$ denotes the smallest absolute difference between two angles $\theta_1,\theta_2$. 

\section{System Model}\label{sec:system}
In this section, we describe the system model used in this paper and define the metrics used for the system evaluation. We also formally define the beamspace representation of a channel. 

\subsection{Antenna model}\label{sec:antenna}
We consider \MU-\MISO \OFDM system. The \BS is equipped with $\Ns$ symbol streams and $\Nrf$ RF chains. It serves $U$ \UEs simultaneously using \MU beamforming. 

In this paper, we only focus on the beam training at the \BS. We assume the \UEs are equipped with omnidirectional antennas with the same polarization as the \BS antenna array. Consequently, each \UE receives a single symbol stream from one \RF chain of the \BS antenna array, resulting in a configuration where $\Ns=\Nrf = U$. Extending this paper to consider \UEs with multiple antennas, hence, a \MIMO system, is a potential area for future research. 

We assume the \BS uses fully connected hybrid antenna architecture with $\Nrf$ \RF chains connected to a uniform rectangular antenna array of $\Nbsx \times \Nbsy$ elements spaced at half-wavelength distance apart. 
The \BS antenna array is positioned on the XY-plane with its broadside direction along the Z-axis~\cite{VaChoShi:Inverse-multipath-fingerprinting:17,Bal:Antenna-Theory:-Analysis:12}. We denote the azimuthal and elevation angles by $\theta \in [-\pi, \pi)$ and $\phi \in [0, \pi]$, respectively, and define $\Omega_x = \cos \theta \sin\phi, \Omega_y = \sin\theta \sin\phi$. Then, the Vandermonde vectors along the $X$ and $Y$ directions can be defined as 
\begin{align}
    \mathbf{a}_{\rm x}(\theta, \phi) & = \frac{1}{\sqrt{\Nbsx}}[1, e^{-\imj \pi\Omega_x}, \ldots, e^{-\imj (\Nbsx - 1) \pi\Omega_x}]^\trans,\label{eq:antenna_response_x}\\
    \mathbf{a}_{\rm y}(\theta, \phi) & = \frac{1}{\sqrt{\Nbsy}}[1, e^{-\imj \pi \Omega_y}, \ldots, e^{-\imj (\Nbsy - 1) \pi \Omega_y}]^\trans.\label{eq:antenna_response_y}
\end{align}
The antenna array response matrix, denoted by $\mathbf{A}(\theta, \phi) \in \bb{C}^{\Nbsy \times \Nbsx}$, can be defined as
\begin{equation}
\mathbf{A}(\theta,\phi) = \mathbf{a}_{\rm y}(\theta, \phi) \mathbf{a}_{\rm x}(\theta, \phi)^\trans.
\end{equation}
Using this antenna model, we now define the channel and signal models. 

\begin{figure*}[tb]
    \centering
    \includegraphics[width=.6\textwidth]{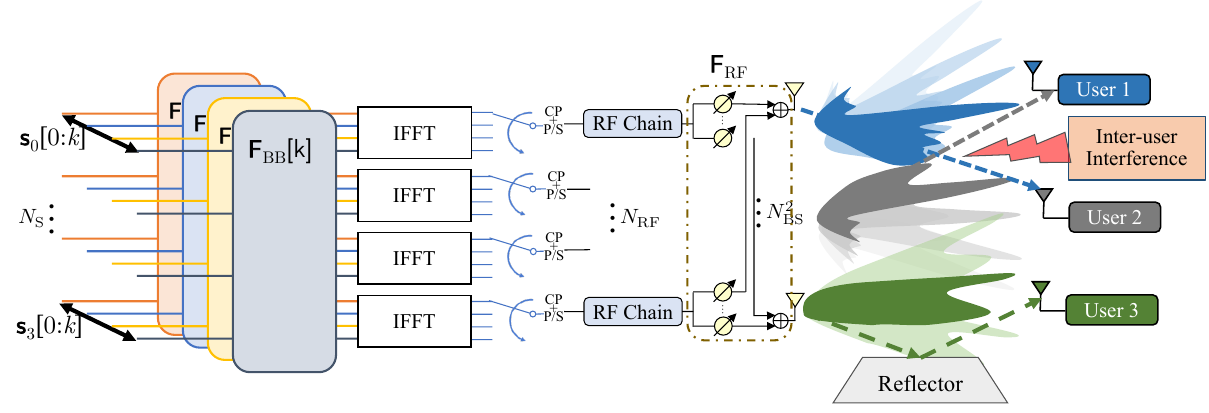}
    \caption{Schematic diagram: We consider a \MU-\MISO \OFDM system with a frequency selective channel model. The \BS equipped with a planar antenna array with hybrid antenna architecture transmits to $U=4$ UEs equipped with an omnidirectional antenna array. Greedily choosing the optimal beam for each UE can introduce significant inter-user interference.}
    \label{fig:scehmatic}
\end{figure*}

\subsection{Channel model}

We consider the wideband channel model with $D$-taps as defined in \cite{GonRodGon:Channel-Estimation-and-Hybrid:18}. We assume the channel is static over one time slot. At a time-slot $n$, let $\left(\alpha_{u,c,\ell}^{(n)},\tau_{u,c,\ell}^{(n)}\right), \forall {\ell \in [L^{(n)}_{u,c}], c\in[C^{(n)}_u]}$ be the gain and delay of the $\ell$-th path in the $c$-th ray cluster to \UE $u$. Let $\left(\theta^{(n)}_{u,c,\ell}, \phi^{(n)}_{u,c,\ell}\right), \forall \ell \in [L^{(n)}_{u,c}],c\in[C^{(n)}_u]$ be the azimuthal and elevation angles of the $\ell$-th path in the $c$-th ray cluster to \UE $u$. 
Further, let $p(t)$ denote the pulse shaping function evaluated at time $t$. 
Consequently, the frequency response of the channel $\bm{\mathsf{H}}^{(n)}_u[k]\in \bb{C}^{\Nbsy \times \Nbsx}$ over the sub-carriers $k\in\{-\lfloor \frac{K-1}{2}\rfloor,\ldots,\lceil\frac{K-1}{2}\rceil\}$ can be defined as
\begin{align}
    \bm{\mathsf{H}}^{(n)}_u[k] & = \sum_{d = 1}^{D}\sum_{c=1}^{C^{(n)}_u} \sum_{\ell = 1}^{L^{(n)}_{u,c}} \left[\alpha_{u,c,\ell}^{(n)} p\left(d\Ts - \tau_{u,c,\ell}^{(n)}\right) \right.\nonumber\\
    & \quad\quad\quad\quad\quad\quad\quad\left.\times \mathbf{A}\left(\theta^{(n)}_{u,c,\ell}, \phi^{(n)}_{u,c,\ell}\right) e^{-\imj \frac{2\pi d k}{K}}\right]. 
\end{align} 
We also denote the vectorized channel $\vectorize{\bm{\mathsf{H}}^{(n)}_u[k]}$ by $\bm{\mathsf{h}}^{(n)}_u[k]$ in the subsequent signal modeling. 

\subsection{Signal model}
We consider a downlink \MU-\MISO system with linear beamforming where each user receives a single data stream. 
Let $\bm{\mathsf{s}}[k] \in \bb{C}^{\Ns \times 1}$ be a vector of symbols such that each element $\bm{\mathsf{s}}_i[k]$ denotes the symbol to a \UE over the $k$-th sub-carrier. We assume $\mathbb{E}[\bm{\mathsf{s}}[k]\bm{\mathsf{s}}^*[k]] = \frac{P}{\Ns}\mathbf{I}_{\Ns}$, where $P$ is the average transmitted power over $k$-th sub-carrier. 

The BS uses hybrid precoding by first digitally precoding each sub-carrier using the digital precoder $\Fbb[k] =[\fbb[1][k], \cdots, \fbb[\Ns][k]] \in \bb{C}^{\Nrf \times \Ns}$, followed by the \RF precoding using the RF precoder $\Frf  =[\frf[1], \cdots, \frf[\Nrf]]\in \bb{C}^{\Nbsx\Nbsy \times \Nrf}$. By denoting the additive Gaussian noise $\mathsf{n}_u[k]\sim \cc{N}(0,\sigma^2)$, the received signal at the intended \UE $u$ can be written as
\begin{align}
    {\mathsf{y}}_u[k] & = \bm{\mathsf{h}}_u^\trans[k]\Frf \fbb[u][k] \bm{\mathsf{s}}_u[k] \nonumber \\
    & + \bm{\mathsf{h}}_u^\trans[k]\Frf \sum_{u'=1, u'\neq u}^{U}\fbb[u'][k] \bm{\mathsf{s}}_{u'}[k] + {\mathsf{n}}_u[k].
\end{align}
Note that the second term in the signal model characterizes the inter-user interference caused by other active users. 
We set the elements of $\Frf$ to have a unit magnitude for modeling the RF precoding by a passive analog phased array. Accordingly, the choice of $\Frf$, specifically, the phase of each element of $\Frf$, defines the beam pattern of the \BS antenna. 

\subsection{Metric: Achievable spectral efficiency (SE)}
With the proposed signal model, we now define the achievable sum \SE of the system to evaluate the performance of the \MU communication system. 
Assuming Gaussian signalling and treating inter-user interference as Gaussian noise, the achievable \SE of the \UE $u$ at a time defined as
\begin{equation}
    R_u = \frac{1}{K}\sum_{k=0}^{K-1} \log_2\left( 1+\frac{\frac{P}{\Ns} \left|\bm{\mathsf{h}}^{\trans}_u[k] \Frf \fbb[u][k]\right|^2 }{\sigma^2+\frac{P}{\Ns}\sum_{u'\neq u} \left|\bm{\mathsf{h}}^{\trans}_u[k] \Frf \fbb[u'][k]\right|^2}\right).
\end{equation}
Further, we define the sum-\SE of the system as $R_\mathrm{S} = \sum_{u=1}^U R_u$.

\subsection{The beamspace of a channel}
We call the list of \AoD pairs of significant paths, $\left(\theta^{(n)}_{u,c,\ell}, \phi^{(n)}_{u,c,\ell}\right),\allowbreak\forall{\ell \in \left[L^{(n)}_{u,c}\right]},\allowbreak c\in C^{(n)}_u$, in the channel the \textit{\AoD-list} of the channel.
Further, we let the 2-D angular space of azimuthal and elevation angles is quantized on the grid of size $G_\theta\times G_\phi$, such that the $(i,j)$-th index of the grid is associated to the direction $(\theta_i,\phi_j),$ where $\theta_i = \frac{2\pi}{G_\theta}i-\pi,\phi_j = \frac{\pi}{G_\phi}j$. Then, $\forall i\in[G_\theta], j\in[G_\phi]$, we define the beamspace representation of the channel as
\begin{align}\label{eq:spatialCSI:PowerProfile}
    \tilde{\mathbf{G}}^{(n)}_{u,(i,j)}=\sum_{c = 1}^{C^{(n)}_u}\Bigg\vert\frac{1}{L^{(n)}_{u,c}}\sum_{\ell = 1}^{L^{(n)}_{u,c}} &\mathbf{a}_{\rm y}^*(\theta_i,\phi_j)\times \nonumber\\
    & \mathbf{A}\left(\theta^{(n)}_{u,c,\ell}, \phi^{(n)}_{u,c,\ell}\right)\times\mathbf{\conj{a}}_{\rm x}(\theta_i,\phi_j)\Bigg\vert^2. 
\end{align}
The proposed beamspace representation of the channel represents the expected strength of the channel along the directions indexed by $(i,j)$ assuming the total gains of all ray clusters are identical. 
This beamspace representation of the channel is a normalized projection of the \textit{physical} channel on the oversampled antenna array manifold~\cite{HeaGonRan:An-Overview-of-Signal-Processing:16}. This projection provides a geometric perspective on the channel as observed by the \BS equipped with planar arrays as defined in \secref{sec:antenna}. 
Consequently, the beamspace representation of a channel also captures the impact of the grating lobes produced by the planar array. For example, if a beam along the direction $(\theta_i,\phi_j)$ produces a grating lobe, and if there exists a path along the direction of the grating lobe, then the beamspace representation captures its impact. 
Furthermore, the beamspace representation does not contain rapidly varying channel features, hence, we characterize it with a frequency flat response.

We acknowledge that there can be a large number of multipath components in \mmWave channels~\cite{SamRap:3-D-Millimeter-Wave-Statistical-Channel:16}. Hence, a complete beamspace of a \mmWave channel can include a lot of multipath components which are infeasible to predict purely from the sensor data. Therefore, we only consider the $\cc{L} (<L^{(n)}_{u,c})$ strongest paths from each ray cluster to define the \textit{truncated} beamspace representation of the channel. 
In this paper, we show that the multimodal sensors in the system can estimate the truncated beamspace representation of the channel (in \secref{sec:stageI}) and the estimated truncated beamspace representation can be used for \MU beamforming in \mmWave communication (in \secref{sec:stageII}). 
Since all discussion for the rest of the paper applies to both – truncated and non-truncated – beamspace representations, we avoid the term truncated for consistency unless explicitly required. 

While the concept of beamspace exists in prior work~\cite{HeaGonRan:An-Overview-of-Signal-Processing:16}, our definition diverges slightly. We assume a path gain of unity for each channel path, and a normalized gain of each multipath cluster because of the difficulty in accurately predicting path gains solely from sensor data. Exploring alternative methods for path gain prediction can be a promising avenue for future research.

Our proposed beamspace representation, focusing solely on \AoDs, aligns with concepts used in prior \mmWave channel estimation techniques. For instance, compressive sensing-based approaches include an intermediate step for identifying the \textit{support set} of the sparse beamspace representations~\cite{RodGonVen:Frequency-Domain-Compressive-Channel:18, GaoDaiHan:Reliable-Beamspace-Channel:17,GaoDaiWan:Channel-estimation-for-mmWave:16,VenAlkGon:Channel-Estimation-for-Hybrid:17}. This support set, defined as the indices of the non-sparse elements in the beamspace representation of the channel, denotes the directions of channel paths without their path gains. Similarly, side-information-aided beamforming solutions use out-of-band channel characteristics to derive a prior over the \AoDs, essentially creating a prior over the beamspace without path gains~\cite{AliGonHea:Millimeter-Wave-Beam-Selection:18,AliHea:Compressed-beam-selection-in-millimeterwave:17,AliGonHea:Spatial-Covariance-Estimation:19}. 
These intermediary steps in prior work can be viewed as a form of beamspace representations without using the path gains. Like these techniques, our algorithm also leverages the proposed AoD-based beamspace as an intermediate step to map sensor data to precoder selection.

\section{Stage I: Sensor-aided beamspace estimate of the channel} \label{sec:stageI}
In this section, we detail the sensor-aided prediction of the channel's beamspace representation. We begin by describing the sensor data preprocessing and the encoding techniques for \AoDs of the channel paths. We then present the \DNN-based multimodal sensor fusion network and loss functions employed for network training.

\subsection{Preprocessing sensor data}
The method of preprocessing the input data (i.e., sensor data) has a great impact on the choice of the \DNN architecture. Thus, in this subsection, we explain the preprocessing steps used in this paper and accordingly, present the \DNN architecture in \secref{sec:NNModel}. 

We consider \UEs equipped with a LiDAR device and a positioning service, while the \BS is equipped with a camera. Given our focus on capturing only the large-scale channel effects for beamforming and assuming a sufficiently low beam retraining period (approximately 500 ms)~\cite{VaChoHea:The-Impact-of-Beamwidth-on-Temporal:15}, we assume the sensors capture samples synchronously with a frequency of at least 10 Hz. We now describe the preprocessing steps for each sensing modality. 

\subsubsection{LiDAR samples}
Each LiDAR sample provides a snapshot of the environment surrounding the UE in the form of a point cloud, where each point represents a reflection of emitted light pulses. Objects introduce a higher density of points due to numerous reflections. Therefore, the number of points in a LiDAR point cloud varies depending on the environment. Such variable-length inputs raise a challenge for designing \DNN models requiring fixed-size inputs. Thus, the key objective of LiDAR data processing is to encode the variable-length sensor data samples into fixed-size samples. 

We adopt the standard LiDAR preprocessing method discussed in \cite{ZhoTuz:VoxelNet:-End-to-End-Learning:18,KlaGonHea:LIDAR-Data-for-Deep:19,SalReuRoy:Deep-Learning-on-Multimodal:22}, which encodes the variable-length sensor data into a fixed-size 3D matrix. 
Specifically, we first define a space corresponding to the BS coverage using the coordinates $(X_{\min}, X_{\max})$, $(Y_{\min},Y_{\max}), (Z_{\min},\allowbreak Z_{\max}).$
We then uniformly partition the 3D space by quantizing each dimension to $b_{\rm x}, b_{\rm y}, b_{\rm z}$ levels, respectively. 
Next, we assign the value $-2$ to the bin containing the \BS and the value $-1$ to the bin containing the \UE. 
Finally, for the remaining bins indexed by $(i,j,k) \in [b_{\rm x}^\cc{L}] \times [b_{\rm y}^\cc{L}] \times [b_{\rm z}^\cc{L}]$, we assign the value $1$ if there exists at least one LiDAR point that falls within the specific bin. 

\subsubsection{Camera images}
The camera captures a visual snapshot of the environment around the \BS in the form of an image. We assume the BS captures a 180-degree \fov using either a single camera or multiple cameras with images stitched together. In this paper, we resize each 180-degree FoV image to a grayscale image of size $b_{\rm h}^\cc{C}\times b_{\rm w}^\cc{C}$ and normalize the pixel values between 0 and 1.

\subsubsection{Coordinates}
We assume that the \UEs have positioning capability from \RF-based localization, \GNSS, sensor-based Simultaneous Localization and Mapping (SLAM)~\cite{Fre:A-discussion-of-simultaneous-localization:06} or other such methods. 
In this paper, we convert the coordinates of \UEs with respect to the \BS location and define a one dimensional input of length~2.

\subsection{Encoding the AoD-list of the channel for training}
In this subsection, we first discuss the motivation for using a novel approach for encoding the \AoD-list of the channel and then propose two encoding techniques. 
The encoded \AoD-list is used as a ``ground truth'' for training the multimodal fusion network described in \secref{sec:NNModel}. Accordingly, we assume that for each sample of the sensor data, the \AoD-list of the associated channel is available during the training process. 

\subsubsection{Motivation}
In this paper, we model the task of predicting \AoD-pairs as a multi-label identification problem~\cite{MadKocGjo:An-Extensive-Experimental-Comparison:12} due to a fundamental difference between our work and the prior work like \cite{KlaGonHea:LIDAR-Data-for-Deep:19,JiaChaAlk:LiDAR-Aided-Future:22,SalReuRoy:Deep-Learning-on-Multimodal:22,WanKlaRib:MmWave-Vehicular-Beam:19}. The prior work have focused on identifying a single beam (or top-$K$ beams) from a codebook based on processed sensor data. Therefore, they framed the problem as a classification task with each beam being a class encoded using one-hot encoding. 
We, however, focus on identifying the \AoD-list of channels with more than one channel paths. Thus, the \AoDs of the channel paths can not be encoded using one-hot encoding. Therefore, we consider a quantized \AoD-pair as a label and treat the \AoD-list prediction problem as the multi-label identification problem with the multi-label encoding for the \AoD-list of the channel. 

In addition to the standard multi-label encoding, we also require an encoding for the \AoD-list that can capture a notion of similarity between channels. For example, a channel path with \AoD $(\theta, \phi) = (0^\circ, 0^\circ)$ is closer to the path with \AoD $(2^\circ, 2^\circ)$ than $(10^\circ, 10^\circ)$. As a result, the channel having only one \AoD $(0^\circ, 0^\circ)$ is closer to the channel having only \AoD $(2^\circ, 2^\circ)$ than the channel having only one \AoD $(10^\circ, 10^\circ)$. 
The conventional solutions to multi-label identification problem and multi-label encoding techniques, does not model this similarity between the ground truths, because of implicit assumption of ``incomparability'' across categorical labels. 

A common approach of leveraging the similarity of the ground truth labels is to model the problem as a regression task. In that case, the \DNN model can take the processed sensor data as an input and predict the directions of all channel paths. A typical \mmWave channel, however, can have a variable number of channel paths, while a fixed size \DNN-based regression model can not handle a variable number of outputs. 

This motivates us to design an encoding technique for the \AoD-list of a channel that \textit{(i)} is an extension of multi-label encoding to allow variable number of \AoDs to be predicted, and \textit{(ii)} can capture the notion of similarity across channels. 

\subsubsection{Encoding techniques}
\begin{figure}[tb]
    \centering
    \includegraphics[width=.8\columnwidth]{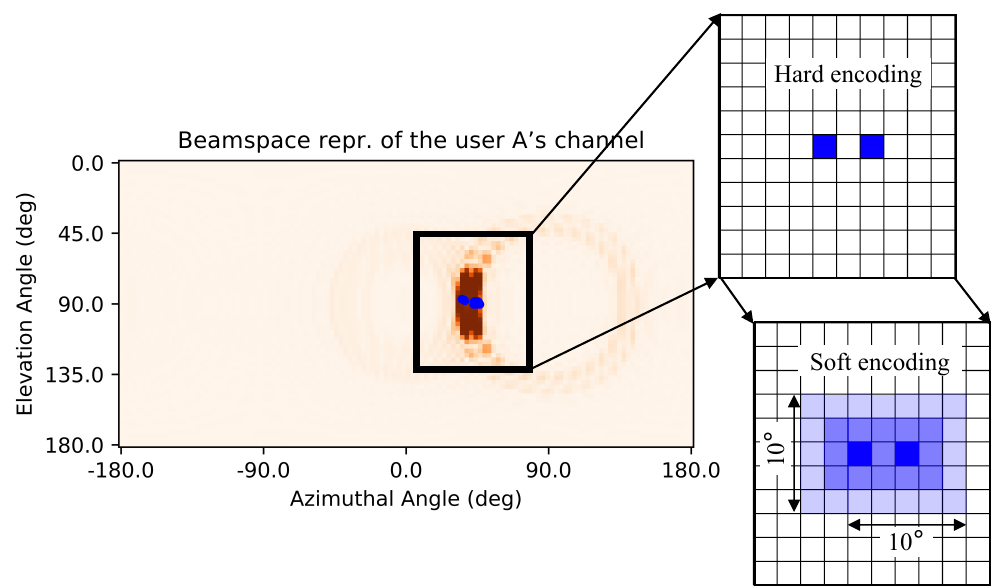}
    \caption{Illustration of hard and soft encoding of the channel's \AoD-list: 
    The blue squares represent a value of 1, indicating the presence of a channel path along those specific directions.
    The soft encoding incorporates a perturbation range of $\Delta = 10^\circ$ in both elevation and azimuthal directions. The value within each square decays linearly from the nearest square containing the actual channel path direction. 
    While the grid is shown for a limited area for illustrative purposes, it would encompass all azimuthal and elevation angle pairs.}
    \label{fig:hard-soft-encoding}
\end{figure}

We define two types of encoding for the list of \AoDs of channel paths: \textit{(a)} hard encoding, \textit{(b)} soft encoding. See \figref{fig:hard-soft-encoding} for a reference.
\paragraph{Hard encoding}
We define the \AoD-pair as a label for the multi-label encoding. Let $(\theta_{i,\ell}, \phi_{i,\ell})$ denote the azimuthal and elevation angles (in degrees) of the $\ell$-th path associated with the channel of sample $i$. Let $Q$ be a quantization map such that $Q(\theta,\phi) = (\quant{\theta},\quant{\phi})$ is an index of $(\theta,\phi)$ quantized on a uniform grid of size $(\Theta,\Phi)$. 
We denote the hard encoding of the \AoD-list associated to the sample $i$ as $\mathbf{y}_i \in \{0,1\}^{\Theta\times\Phi}$ where $\mathbf{y}_i[\quant{\theta}_{i,\ell}, \quant{\phi}_{i,\ell}] = 1, \forall \ell$ such that $Q(\theta_{i,\ell},\phi_{i,\ell}) = (\quant{\theta}_{i,\ell}, \quant{\phi}_{i,\ell})$. 

\paragraph{Soft encoding}
We define the soft encoding as a perturbation on hard encoding (see \figref{fig:hard-soft-encoding}). Formally, considering a perturbation range of $\pm \frac{\Delta}{2}^\circ$ on hard encoding, we define the soft encoding of the \AoD-list associated to the sample $i$ as 
\begin{align}    
\tilde{\mathbf{y}}_i[\quant{\theta}, \quant{\phi}] = 1-\min_{(\ell,\theta,\phi):Q(\theta,\phi) = (\quant{\theta}, \quant{\phi})} & \max\left\{\frac{\angdiff{\theta}{\theta_{i,\ell}}}{\Delta/2},\right.\nonumber\\
& \quad\left.\frac{\angdiff{\phi}{\phi_{i,\ell}}}{\Delta/2}\right\}.
\end{align}
This perturbation adds non-zero, linearly decaying values to angles close to the directions of the channel paths. 

Based on the soft encoding of \AoD-list, we measure the similarity between the channels associated to samples $i,j$ as 
\begin{equation}
    \rho(\tilde{\mathbf{y}}_i, \tilde{\mathbf{y}}_j) = \frac{\tilde{\mathbf{y}}_i^\trans \tilde{\mathbf{y}}_j}{||\tilde{\mathbf{y}}_i||_2 ||\tilde{\mathbf{y}}_j||_2}. \label{eq:similarity-metric}
\end{equation}
The proposed metric measures the overlap of channel path directions within the defined perturbation range. Higher overlap in the soft encoding of the \AoD-list of two samples within the perturbation range translates to a higher similarity score. Additionally, the normalization in the metric ensures that both channels are weighted equally regardless of the number of paths they contain. This also guarantees that if one channel encoding $\tilde{\mathbf{y}}_i$ has more paths, the other channel $\tilde{\mathbf{y}}_j$ would require a higher number of paths in similar directions to get considered as highly similar. 

The hard and soft encodings capture different aspects of the channel's \AoD-list. Hard encoding defines a clear boundary between valid and invalid channel directions for a specific sample, while the soft encoding incorporates the notion of similarity between different channels even if they are not identical. In this paper, we leverage both these aspects in DNN training, employing different loss functions for each encoding, as detailed in \secref{sec:nn-metrics}. 

We can further extend the proposed encoding methods beyond the current definition of the index for hard and soft encodings, which represent angle pairs $(\theta, \phi)$ (referred to as angle-based encodings). As an alternative, we can define hard- and soft-encoded vectors indexed by quantized triplets of cosines $(\sin(\theta), \cos(\theta), \sin(\phi))$ (cosine-based encodings). This approach introduces non-uniform quantization of angles compared to the uniform quantization used in angle-based encodings. Intuitively, cosine-based encoding should be a better choice for training since precoder design depends on the cosines of angles and not the exact angles themselves. Accordingly, we evaluate the performance of both angle-based and cosine-based encodings during the training and testing phases of the \DNN model.

Note that the soft encoding in itself is not enough for the \DNN model to learn the similarity/dis-similarity between channels. We also require an appropriate loss function for training \DNN, which we propose in \secref{sec:nn-metrics}.

\subsection{Multimodal sensor fusion: Design of DNN}\label{sec:NNModel}
In this subsection, we describe the \DNN-based multimodal sensor fusion model, which leverages the pre-processed sensor data and hard- and soft-encoded \AoD-lists of the associated channels. We first outline the schematic of our model, containing unimodal feature extractors and fusion network. We then describe the two-part structure of the fusion network -- the key novelty of our model. 

\begin{figure}[tb]
    \centering
    \includegraphics[width=.8\columnwidth]{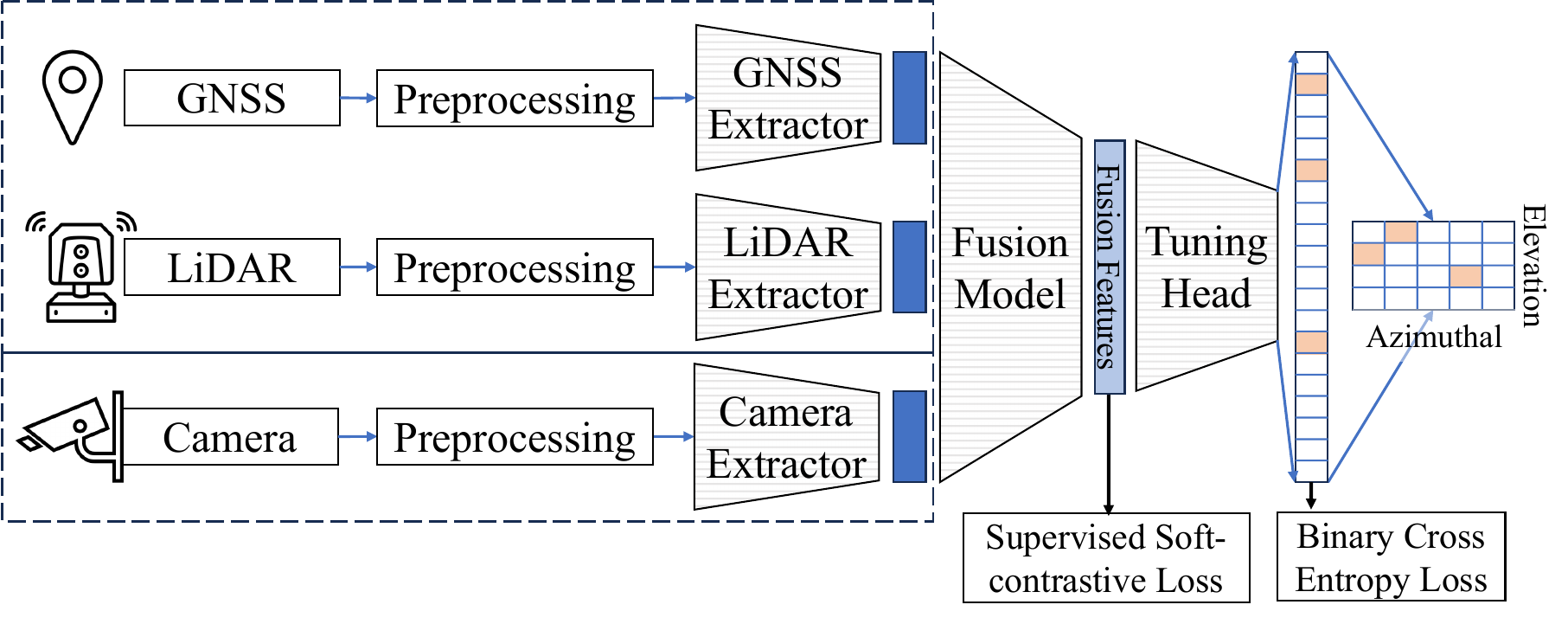}
    \caption{Schematic diagram of the proposed \DNN-based multimodal fusion network}
    \label{fig:nn-model-schematic}
\end{figure}

\begin{figure*}[tb]
    \centering
    \includegraphics[width=0.7\textwidth]{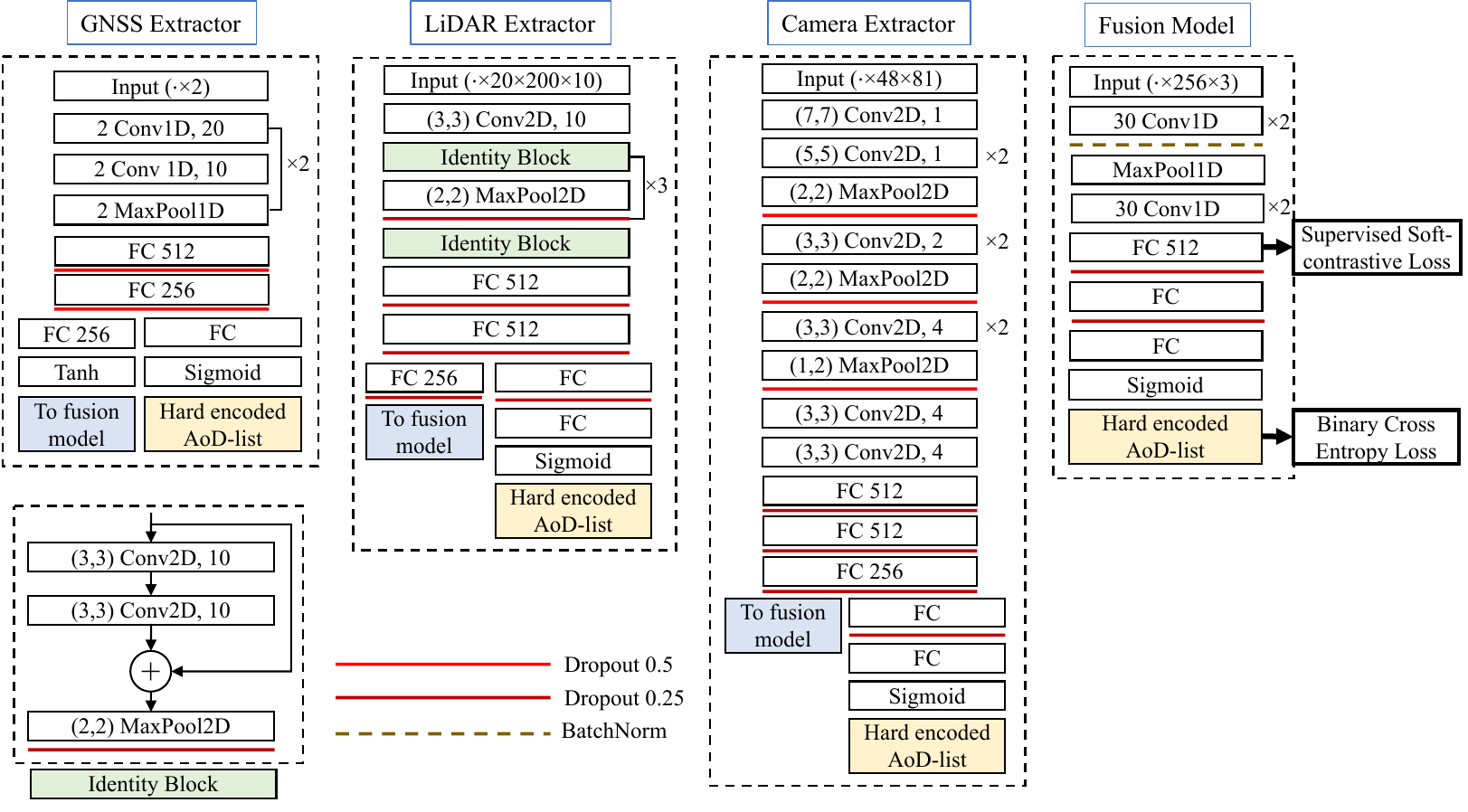}
    \caption{Detailed block diagrams of unimodal feature extractors and the fusion network: The branches of the unimodal extractors are used depending on whether the extractors are used with the fusion model or to directly predict the \AoD-list of a channel.}
    \label{fig:nn-model-detailed}
\end{figure*}

Our model draws inspiration from \cite{SalReuRoy:Deep-Learning-on-Multimodal:22}. We, however, modify the hyper-parameters near the network's head to accommodate the larger output dimensions required for our task. 
Similar to \cite{SalReuRoy:Deep-Learning-on-Multimodal:22}, our model operates in two stages as illustrated in \figref{fig:nn-model-schematic}: 
\begin{enumerate}
    \item The pre-processed data from each modality is passed through its corresponding \textit{unimodal feature extractor} (detailed in \figref{fig:nn-model-detailed}). These extractors generate modality-specific features. 
    \item The features extracted from each modality are then concatenated and fed as the input to the \textit{fusion network} (detailed in \figref{fig:nn-model-detailed}). 
\end{enumerate}

The key novelty in our model is the two-part structure of the fusion network. 
Specifically, the first part, called \textit{fusion feature extractor}, takes the concatenated features as an input and generates \textit{fusion features} of dimension $512$. We denote the fusion features associated with the $i$-th sensor data sample by $\mathbf{\hat z}_i$. These fusion features are used to incentivize the unimodal feature extractors and the fusion network to output similar fusion features for similar channels. We describe the incentive in \secref{sec:nn-metrics}. 

The second part of the fusion network, called \textit{tuning head}, uses the fusion features $\mathbf{\hat z}_i$ to predict the hard-encoded \AoD-list $\mathbf{y}_i$ in the form of probabilities denoted by $\mathbf{\hat{y}}_i$. We consider a value greater than $0.5$ in $\mathbf{\hat{y}}_i$ as a positive prediction. 
Finally, for each sample $i$ associated to \UE $u$ and time-slot $n$, the estimated $\mathbf{\hat y}_i$ is used to obtain the beamspace representation of the channel $\hat{\tilde{\mathbf{G}}}_u^{(n)}$. 

\subsection{Loss functions and evaluation metrics}\label{sec:nn-metrics}
In this subsection, we describe the loss functions used to train the entire multimodal fusion model, and the metrics used to evaluate its performance.

\subsubsection{Binary cross entropy (BCE) loss}
BCE loss is used during training to minimize the disparity between the predicted AoD-list $\mathbf{\hat{y}}_i$ and the ground truth represented by the hard-encoded AoD-list $\mathbf{y}_i$. Mathematically, for a batch of sensor observations denoted by $\cc{B}$, the BCE loss is defined as 
\begin{equation}
    \textrm{BCE}(\mathbf{y},\mathbf{\hat{y}}) = -\frac{1}{|\cc{B}|}\sum_{i \in \cc{B}} \left(\mathbf{y}_i^\trans \log(\mathbf{\hat{y}}_i) + (\mathbf{1}-\mathbf{y}_i^\trans) \log(1-\mathbf{\hat{y}}_i)\right).
\end{equation}

\subsubsection{Supervised soft-contrastive loss (SSCL)}
\SSCL is used during training to incentivize the \DNN to learn the notion of similarity between channels by encouraging to extract similar features from the fusion of sensor data for similar channels. For that, we introduce two objectives on the fusion features $\mathbf{\hat z}_{i}$ and $\mathbf{\hat{z}}_j$ associated to sensor observations $i$ and $j$: \textit{(i)} Aligning the fusion features $\mathbf{\hat z}_{i}$ and $\mathbf{\hat z}_{j}$ when the corresponding sensor observations $i$ and $j$ have similar channels, and \textit{(ii)} dis-aligning the fusion features $\mathbf{\hat z}_{i}$ and $\mathbf{\hat z}_{j}$ when the observations exhibit considerably different channels. This allows robust training of \DNN against the noise in the sensor observations, and further reducing the amount of data required for the training~\cite{KhoTetWan:Supervised-contrastive-learning:20}.

To formally define \SSCL, recall the similarity metric $\rho_{i,j}:=\rho(\tilde{\mathbf{y}}_i, \tilde{\mathbf{y}}_j)$ defined in \eqref{eq:similarity-metric}. Then, for a batch of sensor observations $\cc{B}$, we define \SSCL as 
\begin{align}
    \textrm{SSCL}(\mathbf{\hat z},\mathbf{\tilde y}) & = \frac{1}{|\cc{B}|^2} \sum_{i,j}\left[ -\rho_{i,j} \exp\left(\frac{\mathbf{\hat z}_i^\trans \mathbf{\hat z}_j}{||\mathbf{\hat z}_i||_2||\mathbf{\hat z}_j||_2}\right) \right.\nonumber\\
   &\quad\quad\quad\quad\left. + (1-\rho_{i,j})\exp\left(\frac{\mathbf{\hat z}_i^\trans \mathbf{\hat z}_j}{||\mathbf{\hat z}_i||_2||\mathbf{\hat z}_j||_2}\right)\right], \label{eq:alignment-disalignment-scl} \\
    & = \frac{1}{|\cc{B}|^2} \sum_{i,j} (1-2\rho_{i,j})\exp\left(\frac{\mathbf{\hat z}_i^\trans \mathbf{\hat z}_j}{||\mathbf{\hat z}_i||_2||\mathbf{\hat z}_j||_2}\right). \label{eq:scl}
\end{align}
The first and second part of \eqref{eq:alignment-disalignment-scl} incentivize the fusion model to align and dis-align the fusion features depending on the similarity and dis-similarity of beamspaces in the batch. 
Not only that, the incentives for alignment and dis-alignments are proportional to the similarity and dissimilarity metrics $\rho_{i,j}$, $(1-\rho_{i,j})$, respectively. Thus, \SSCL proposed in \eqref{eq:scl} balances the trade-offs between alignment and dis-alignment of fusion features. 

The proposed \SSCL differs from \SCL presented in \cite{KhoTetWan:Supervised-contrastive-learning:20} in how similarity is weighted. Our approach leverages a continuous metric $\rho(\tilde{\mathbf{y}}_i, \tilde{\mathbf{y}}_j)$ to weigh the alignment/disalignment of fusion features, whereas the original \SCL relies on a binary similarity based on class labels. This fundamental difference necessitates modifications to the original \SCL formulation. Due to the continuous similarity metric as a weight, the multiplicative dis-alignment factor in the denominator of \SCL in \cite{KhoTetWan:Supervised-contrastive-learning:20} can not be ensured to be normalized. Thus, we use an additive dis-alignment factor as in \eqref{eq:alignment-disalignment-scl} instead of a multiplicative factor as in \SCL~\cite{KhoTetWan:Supervised-contrastive-learning:20}. 

Finally, based on the ground truth encodings $(\mathbf{y}, \tilde{\mathbf{y}})$, we use the sum of $\textrm{BCE}(\mathbf{\hat{y}}, \mathbf{y})$ and $\textrm{SSCL}(\mathbf{\hat z},\mathbf{\tilde y})$, as the loss function to train the \DNN. 

\subsubsection{Mean angular distance (MAD) metric}
MAD is used during evaluation to quantify the average discrepancy between the predicted and actual \AoDs. To calculate the MAD, we follow four steps: First, for each predicted \AoD pair, we identify the nearest \AoD pair in the ground truth \AoD-list. Second, we compute the angular distance between the true and predicted \AoD pairs. The angular distance is defined as the inverse cosine of cosine distance between the unit vectors along predicted and true \AoD pairs. 
Third, we average the measured distances across all predicted \AoDs within a sample. Finally, we calculate the average of the sample-wise MAD across all samples in a batch. 

\subsubsection{Mean absolute error (MAE) in cosines metric}
Since the antenna array response vectors depend on cosine-triplet of AoD-pairs, defined as $(\sin(\theta),\cos(\theta),\sin(\phi))$, MAE in cosines is used during evaluation to quantify the discrepancy between the predicted and actual cosine-triplets. 
The calculation of MAE in cosines is similar to that of MAD; however instead of calculating the angular distance between \AoD pairs, we calculate the $L_1$-distance between the cosine-triplets of predicted and true \AoD pairs. 

\begin{remark}
    The number of paths can vary across different samples. Consequently, MAD and MAE-based metrics, which depend on predicted angles, are not suitable as loss functions for \DNN training due to challenges with gradient-based backpropagation. Nevertheless, they remain valuable for assessing the \DNN's performance in terms relevant to the practical application. 
\end{remark}

With the specified design of the network and the loss functions, we have the first stage of the sensor-aided MU beamforming: A model to predict the beamspace representation $\hat{\tilde{\mathbf{G}}}_u$ from sensor observations. We report the performance of training and testing with various sensor modalities in \secref{sec:results:stageI}. In the following, we move on to the second stage, using the predicted beamspaces for \MU beamforming strategy. 

\section{Stage II: Beamspace-based MU beamforming algorithm} \label{sec:stageII}

In this section, we describe the \MU beamforming algorithm that uses the beamspace representations of the users' channels. Given the estimate of the beamspace from multimodal sensor fusion network, we first introduce the intuition behind \MU beamforming algorithm followed by the outline.

\subsection{Key intuition behind the MU beamforming algorithm}

\begin{figure}[tb]
    \centering
    \includegraphics[width=0.3\textwidth]{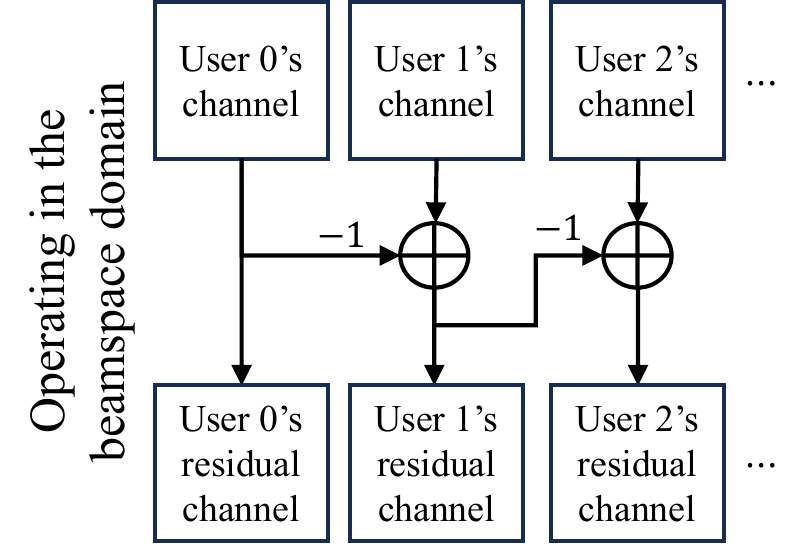}
    \caption{Block diagram for determining residual channels of users based on beamspace estimates: The \BS iteratively removes the channel paths that overlap with other \UEs to get the residual channels. These residual channels of \UEs are used to find the RF precoder $\Frf$.}
    \label{fig:algorithm-flow-diagram}
\end{figure}

\begin{figure*}
    \centering
    \subfloat[Beamspace representation of user ${A}$'s channel\label{fig:intuition-channels-A}]{\includegraphics[width=.3\textwidth]{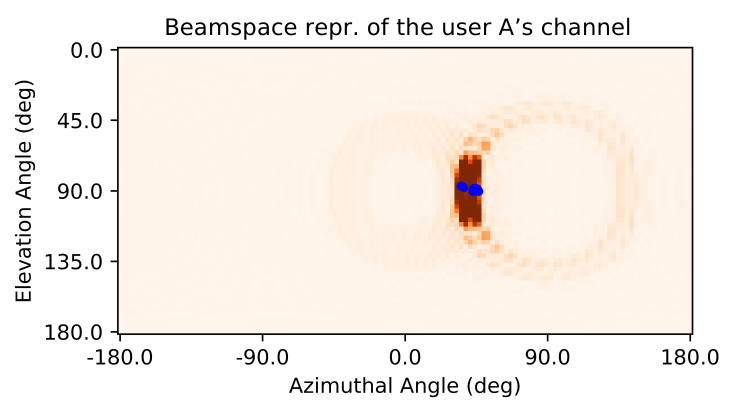}}\quad
    \subfloat[Beamspace representation of user ${B}$'s channel\label{fig:intuition-channels-B}]{\includegraphics[width=.3\textwidth]{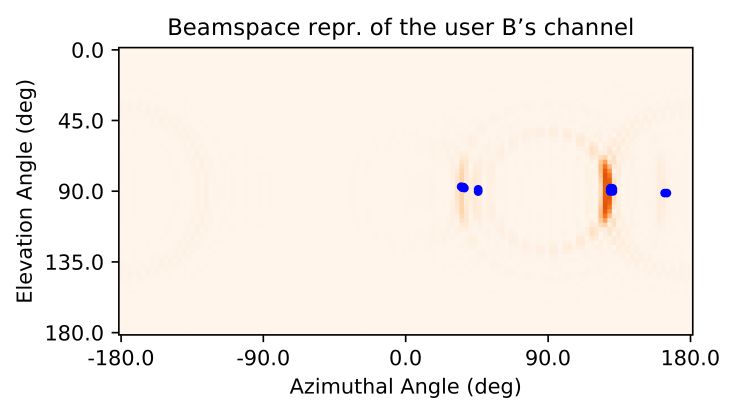}}\\
    \subfloat[Beamspace representation of user ${A}$'s residual channel\label{fig:intuition-effective-A}]{\includegraphics[width=.3\textwidth]{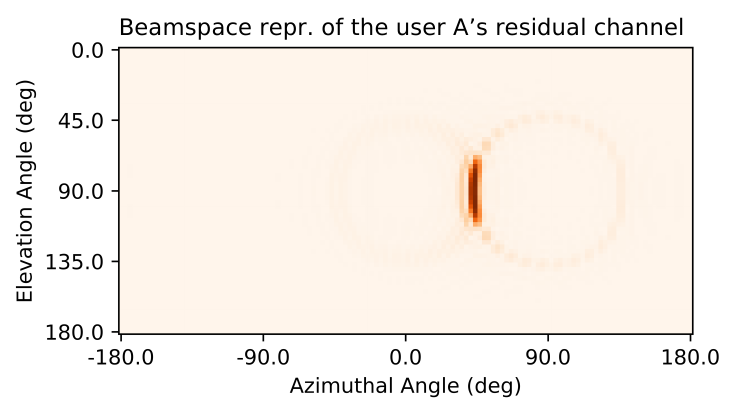}}\quad
    \subfloat[Beamspace representation of user ${B}$'s residual channel\label{fig:intuition-effective-B}]{\includegraphics[width=.3\textwidth]{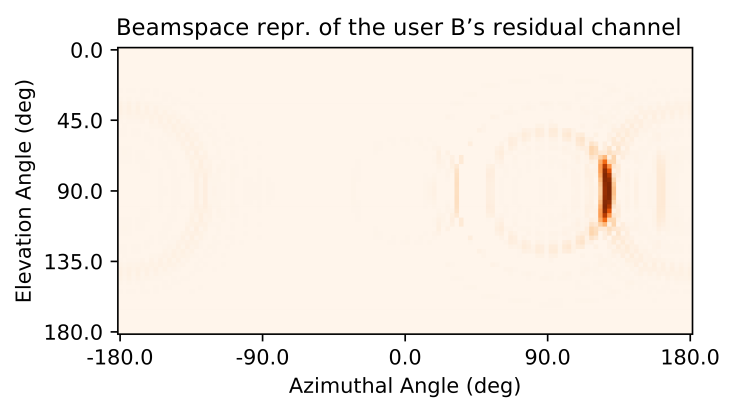}}
    \caption{Intuition behind \MU beamforming algorithm: The overlapping directions in the beamspace representation of user $A$ and user $B$'s channels are removed from the beamspace of user $B$. User $A$ gets the priority because it has fewer \AoDs. The resultant beamspace representations are called residual channels and are used to determine the initial RF precoder $\Frf$.}
    \label{fig:intuition}
\end{figure*}

The algorithm is split into two phases. The first phase focuses on designing \RF precoder from the beamspace representations (see \figref{fig:algorithm-flow-diagram}). The key idea in this phase is to assign one transmission direction to at most one \UE. For instance, consider a scenario where the \BS communicates with two \UEs, $A, B$. The beamspaces of their channels are shown in \figref{fig:intuition-channels-A} and \ref{fig:intuition-channels-B}, respectively. Notice that both \UEs have an overlapping multipath component at \AoD-pair $(45^\circ,90^\circ)$. The \BS starts with the \UE having the least number of predicted paths (in this case, \UE $A$), selects one direction from the its estimated beamspace and removes that direction from the beamspaces of the subsequent users. In this case, for the \UE ${A}$, the algorithm selects the (only) direction from its beamspace and removes that direction from the beamspace of \UE ${B}$. We call the resultant beamspaces, the residual channels of \UEs. The \BS then designs the \RF precoders, $\frf[A],\frf[B]$, that maximizes the transmission energy along the peak of the \UEs' residual channels. This process allows for assigning only one \UE to each cluster of the channel.

In the second phase of the algorithm, the \BS optimizes the digital precoder $\Fbb$ using conventional \MU-\MISO strategies based on the estimation of equivalent channel and channel feedback. Since the dimension of $\Fbb$ is significantly smaller than $\Nbsx \times \Nbsy$, this phase contributes minimally to the overall overhead. 

\subsection{Algorithm outline}
The pseudocode of the algorithm is presented in \algref{alg:main}. 
Consider the $U = \Ns = \Nrf$ number of active \UEs. 
In the first phase, the \BS receives the predicted \AoD-list and the associated beamspace estimates $\hat{\tilde{\mathbf{G}}}_u$ of all \UEs from the fusion network. The \BS then sorts \UEs in the ascending order of the number of predicted paths. Then for each \UE $i$, the \BS creates the residual beamspace representation, denoted by $\check{\mathbf{G}}_i$, by subtracting the residual beamspaces of all \UEs $j < i$. 
It then identifies the directions $(\hat{\theta}_i, \hat{\phi}_i)$ associated with the peak of $\check{\mathbf{G}}_i$, and selects $\frf[i] = \vectorize{\conj{\mathbf{A}}(\hat{\theta}_i, \hat{\phi}_i)}$ as the RF precoder for \UE $i$. This process is repeated until all \UEs are assigned an \RF precoder. 

\begin{algorithm}[tb]
    \begin{algorithmic}[1]
        \STATE \textbf{Input:} An ordered list of \UEs $\cc{U} = (u_1, \ldots, u_U)$ to be communicated, sorted in the ascending order of the number of the predicted paths.
        \STATE \textbf{Input:} Beamspace estimate $\hat{\tilde{\mathbf{G}}}_u, \forall u \in \cc{U}$
        \STATE \textbf{First stage: } Using beamspace estimate to predict $\Frf$
        \FOR{$i = 1, 2, \ldots, U$}
            \STATE Residual beamspace: $\check{\mathbf{G}}_i = \hat{\tilde{\mathbf{G}}}_i - \sum_{j = 0}^{i-1} \hat{\tilde{\mathbf{G}}}_j $
            \STATE Find a peak direction in $\check{\mathbf{G}}_i$, denoted by $(\hat\theta_i, \hat\phi_i)$.
            \STATE Design a directional beamformer $\frf[i]= \vectorize{\conj{\mathbf{A}}(\hat{\theta}_i, \hat{\phi}_i)}$.
        \ENDFOR
        \STATE Set $\Frf = [\frf[1], \cdots, \frf[\Nrf]]$.
        \STATE \textbf{Second stage: } Designing \MU digital precoder
        \STATE \BS sets $\Fbb[k] = \mathbf{I}_{\Nrf}$ and transmit training sequence to all \UEs simultaneously. 
        \STATE All \UEs estimate their effective channels $\hat{{\bm{\mathsf{h}}}}^\trans_u[k] = \bm{\mathsf{h}}_u^\trans[k]\Frf $ and feeds back to \BS. 
        \STATE Using the estimated channels, the \BS designs \RZF digital precoder $\Fbb[k]$.
    \end{algorithmic}
    \caption{\MU beamforming using the beamspace representations}
    \label{alg:main}
\end{algorithm}

Once $\Frf$ is determined, the \BS then optimizes $\Fbb$ in the second phase. 
This process is similar to the procedure used in preamble-based \MU-MIMO channel estimation~\cite{AlkLeuHea:Limited-Feedback-Hybrid:15, HeaLoz:Foundations-of-MIMO-Communication:18,DreHea:Machine-Learning-Codebook:23}. The \BS transmits a common preamble sequence to all \UEs. 
Each \UE estimates its effective channel $\hat{{\bm{\mathsf{h}}}}^\trans_u[k] = \bm{\mathsf{h}}_u^\trans[k]\Frf$, and feeds it back to the \BS through the control channel. Finally, the \BS collects the estimated channels from all \UEs, and designs the \RZF digital precoder $\Fbb[k]$ as follows
\begin{equation}
    \Fbb[k] = \frac{\left(\sum_{u'=1}^U \conj{\hat{\bm{\mathsf{h}}}}_{u'}[k]\hat{{\bm{\mathsf{h}}}}^\trans_{u'}[k] + \mathbf{I}_{\Nrf}\right)^{-1}\conj{\hat{\bm{\mathsf{h}}}}_u[k]}{\left\Vert\left(\sum_{u'=1}^U \conj{\hat{\bm{\mathsf{h}}}}_{u'}[k]\hat{{\bm{\mathsf{h}}}}^\trans_{u'}[k] + \mathbf{I}_{\Nrf}\right)^{-1}\conj{\hat{\bm{\mathsf{h}}}}_u[k]\right\Vert_2}.
\end{equation}
With the chosen precoders $\Frf, \Fbb[k],\forall k\in[K]$, the \BS can transmit to all \UEs simultaneously. 

\section{Numerical results}\label{sec:results}
In this section, we present numerical evaluation of the proposed sensor-aided \MU beamforming strategy. We first describe the dataset used for evaluation, followed by a comparison of the achieved \MU \SE $R_\text{S}$ with the baselines. We then discuss the impact of using different sensing modalities on the performance and compare the performance of \MU communication with the proposed beamforming strategy to a time-division multiplexed \SU strategy. Finally, we quantify the reduction in the link establishment overhead due to the sensor-aided beamforming approach.

\subsection{Dataset}
To evaluate the performance of the sensor-aided beamspace prediction model, we use the Raymobtime dataset~\cite{KlaBatGon:5G-MIMO-Data-for-Machine:18}. This dataset comprises a collection of synthetic ray tracing datasets designed for simulating wireless channels. We specifically employ datasets \dtrain and \dtest. These datasets offer a rich combination of multimodal sensing data, including LiDAR, camera images and coordinates, alongside ray tracing-based channels at 60 GHz for all ten mobile receivers, in a synthetic 3D scenario of the city Rosslyn, Virginia. 
These datasets capture 2086 and 2000 scenes, respectively, with each scene separated by 30 seconds. 

Crucially, \dtrain and \dtest datasets include the underlying \mmWave channel associated with each sensing data sample. This valuable feature allows training and testing of the proposed prediction model using the sensor data as input and the hard- and soft-encoded beamspace estimates of the underlying mmWave channels as output. In contrast, other real-world multimodal sensing datasets such as DeepSense 6G~\cite{AlkChaOsm:DeepSense-6G:-A-Large-Scale-Real-World:23} and e-FLASH~\cite{SalReuRoy:Deep-Learning-on-Multimodal:22}, used for \mmWave beam predictions, only contain the mapping of the beams in the codebook and measurements of the received powers, but not the complete \CSI. This limits their use in our work. 

Similar to previous work \cite{KlaGonHea:LIDAR-Data-for-Deep:19,SalReuRoy:Deep-Learning-on-Multimodal:22}, we use datasets \dtrain for training our fusion model, while \dtest is used for validating the obtained results. 

\subsection{System setup and parameters for DNN}
Our system uses a \BS equipped with $2\times 1$ \RF chains with $16\times 8$ antenna elements for $U=2$ and $2\times 2$ \RF chains with $16\times 16$ antenna elements for $U=4$. The granularity for the beamspace representation is $G_\theta \times G_\phi = 64\times 32$ for $U=2$ and $G_\theta \times G_\phi = 64\times 64$ for $U=4$. We consider total of $\cc{L} = 25$ strongest paths from all ray clusters combined to define the \textit{truncated} beamspace of the channel. 
Following 5G NR specifications, we use OFDM symbols with $792$ subcarriers spaced at $120$ kHz~\cite{3GP:User-Equipment-UE-radio:22}. We consider all users transmit simultaneously on all subcarriers, and do not consider resource allocation over the resource grids. 

For LiDAR point clouds preprocessing, we define the coverage space of \BS as $(X_{\rm min}, X_{\rm max}) = (744\ {\rm m},767\ {\rm m})$, $(Y_{\rm min}, Y_{\rm max}) = (429\ {\rm m},679\ {\rm m})$, $(Z_{\rm min}, Z_{\rm max}) = (0\ {\rm m},10\ {\rm m})$. Any points outside this range are removed from the point cloud. The remaining data is then quantized into the grid of $b_{\rm x}^\cc{L}\times b_{\rm y}^\cc{L}\times b_{\rm z}^\cc{L} = 20\times 200\times 10$ bins in $X,Y$ and $Z$ directions. The camera images are scaled to the size $b_{\rm h}^\cc{C}\times b_{\rm w}^\cc{C} = 48\times 81$. 

Finally, for the angle-based encoding of beamspaces, we uniformly quantize the azimuthal and elevation angle-pairs into $\Theta \times \Phi = 90 \times 45$ bins. For cosine-based encoding, we uniformly quantize the values of $(\sin \theta, \cos \theta, \sin \phi)$ in $40\times40\times40$ bins. Unless otherwise mentioned, we use angle-based encoding for the most part of the subsequent analysis. For soft-encoding, we consider the perturbation range of $\Delta = 10$.

For the training of DNN model, we use both BCE and SSCL losses with equal weights. We use the batch size of 32. We consider initial learning rate of 0.0001 and reduce it by a factor of 0.99 when the total training loss has stopped improving for 10 consecutive epochs. 

\subsection{MU spectral efficiency}\label{sec:results:stageII}
We evaluate the performance of our complete framework on \dtest dataset. We consider simultaneous transmission to two or four mobile \UEs out of the ten available in each episode. For each episode, we consider all possible combinations of \UEs called \textit{user-clusters}. For each cluster, we use all available sensing modalities -- camera images, LiDAR point clouds and coordinates of the \UEs\ -- to predict the beamspace of the channel. We then use the \MU beamforming algorithm to determine the \RF precoder $\Frf$ and digital precoder $\Fbb$ and calculate \SE $R_u$ achieved by each user $u$ in the user-cluster. 
We then select the user-clusters that provides the maximum \SE for each user. 
Finally, we plot the median and the 25th and 75th percentile values of the sum-\SE of all such user-clusters across all episodes. 

We compare our strategy against three baselines:
\begin{itemize}
    \item Full \CSI at the \BS: In this scenario, the \BS designs $\Fbb$ assuming noise-less channel estimate and a fully digital antenna array architecture. This represents an best case scenario for a \MU-MIMO communication system. 
    \item Beam prediction: The \BS uses the optimal beam for each \UE individually from a DFT codebook. This baseline represents the best case performance of the state-of-the-art multimodal sensor-aided \mmWave beam training framework~\cite{SalReuRoy:Deep-Learning-on-Multimodal:22} when extended to \MU setting without prediction error. 
    \item Ground truth beamspace: The \BS has access to the true truncated beamspace representation, as opposed to the sensor-aided estimate. This baseline demonstrates the effectiveness of our \MU beamforming algorithm, isolating the impact of errors in the sensor-aided beamspace estimates.
\end{itemize}
It is important to note that noise affects the design of $\Fbb$ in our framework and the baselines except in the full \CSI setting, as it requires estimating the channel using preamble-based \MU-MIMO channel estimation method. 
We also emphasize here that we do not consider the overhead of searching the optimal beam from the top-$K$ prediction, neither we consider the overhead of channel estimation. Our focus for the analysis is to highlight the sum-\SE performance of the proposed strategy compared to baseline. Therefore, we assume the access to an ideal beam predictor that selects the best beam and the access to noisy channel estimate without any overhead. We compare the overhead of the proposed strategy with baselines in \secref{sec:results:overhead}. 

\begin{figure}[tb]    \centering
    \subfloat[$U=2$\label{fig:rate-power-tuned-u2}]{\includegraphics[width=0.49\columnwidth]{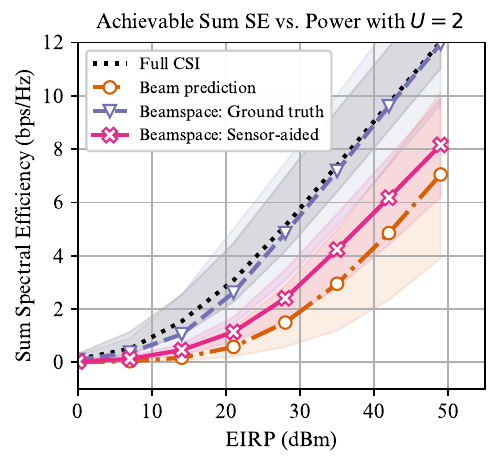}}
    \subfloat[$U=4$\label{fig:rate-power-tuned-u4}]{\includegraphics[width=0.49\columnwidth]{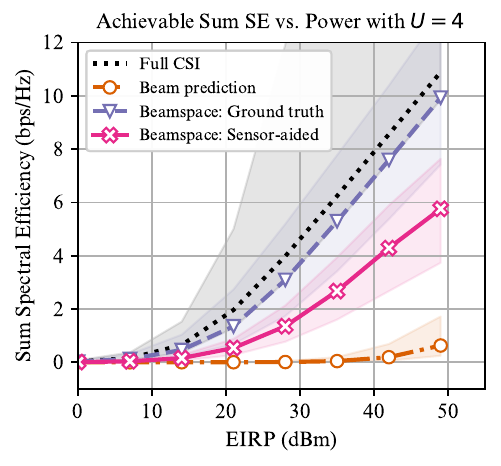}}
    \caption{\MU sum-\SE vs. power for $U=2$ and $U=4$: The lines in the plots denote the median sum-\SE while the shaded regions show the respective 25\%-tile to 75\%-tile range.}
    \label{fig:result-rate-power-tuned}
\end{figure}

\begin{figure}[tb]
    \centering
    \subfloat[$U=2$\label{fig:rate-power-tuned-u2-hist}]{\includegraphics[width=0.49\columnwidth]{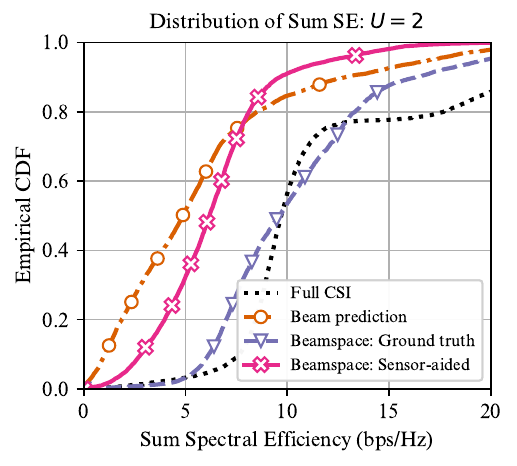}}
    \subfloat[$U=4$\label{fig:rate-power-tuned-u4-hist}]{\includegraphics[width=0.49\columnwidth]{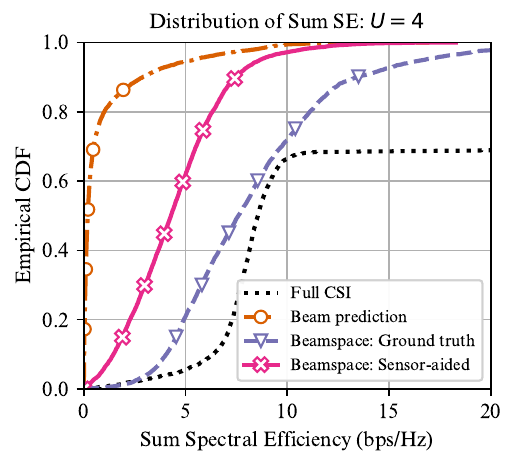}}
    \caption{Distribution of \MU sum-\SE for $U=2$ and $U=4$: beamspace-aided beamforming achieves much higher sum-\SE compared to the beam prediction baseline at 42 dBm \EIRP.}
    \label{fig:result-rate-power-tuned-hist}
\end{figure}

\figref{fig:result-rate-power-tuned} shows the median sum-\SE with 25\% and 75\%-tile range achieved in the \MU setting with both $\Frf$ and $\Fbb$ derived using \algref{alg:main}. This demonstrates the end-to-end performance of combining our proposed \DNN-based multimodal fusion network with \MU beamforming strategy. We observe consistently high performance using beamspace-aided \MU beamforming compared to the \MU extensions of \SU baseline technique. Note that the performance of the beam prediction baseline is significantly worse with high number of active users, indicating the impact of inter-user interference. 
In \figref{fig:result-rate-power-tuned-hist}, we show the empirical CDF of achieved sum-\SE at 42 dBm \EIRP. We observe that the median \MU sum-\SE of the proposed \MU beamforming strategy, specifically with $U=4$, is significantly higher (upto 4$\times$ with $U=4$) than the \MU extension of \SU baseline. 

Note that the prediction error in sensor-aided beamspace estimate, defined in comparison with ground truth beamspace, only depends on the sensor modalities used during training, and remains constant across different number of \UEs. 
From \figref{fig:result-rate-power-tuned-hist}, we observe that there is a larger impact on performance for the same prediction error when using 4 \UEs compared to 2 \UEs. 
This suggests that the proposed algorithm grows more sensitive towards the errors in beamspace prediction as the number of users increases.

Interestingly, in the ideal scenario with known ground truth beamspace, the system performance is close to that of a digital \MU-MIMO system with full CSI setting. This, however, is not the case when using sensor-aided beamspace estimate. The prediction error can introduce mis-alignments or identify sub-optimal directions of multipath components in the channel. Because this prediction error is independent of transmit power, the resultant loss of dominant directions has proportionally high impact for higher transmit power. Therefore, we believe that further optimization of the multimodal fusion network has the potential to improve \SE performance of the proposed \MU framework.

Finally, we also observe an interesting feature of the Raymobtime dataset from the performance of full CSI in \figref{fig:result-rate-power-tuned-hist}. 
We see two distinct groups of user clusters. The top 20\% achieve a sum \SE greater than 18 bps/Hz, while the bottom 60\% have a sum-\SE below 9 bps/Hz. 
Since we only focus on user clusters offering the best individual \SE to at least one user, this sharp difference suggests many users have inherently low SE regardless of other users in their cluster. We believe this low \SE stems from similar \AoDs in the channels of these users. This similarity results in severe inter-user interference, even with a fully digital antenna system.

\begin{table*}[tb]
\centering
\begin{tabular}{r||c|c||c|c}
  & \multicolumn{2}{c||}{\textbf{With angle-based encoding}} & \multicolumn{2}{c}{\textbf{With cosine-based encoding}} \\ \cline{1-5}
 \textbf{Sensor modalities} & \textbf{MAD (deg)} & \begin{tabular}[c]{@{}c@{}}\textbf{MAE in }\\\textbf{ cosine}\end{tabular} & \textbf{MAD (deg)} & \begin{tabular}[c]{@{}c@{}}\textbf{MAE in }\\\textbf{ cosine}\end{tabular} \\ \hline\hline
\textbf{Coordinates} & 7.2821 & 0.1258 & 15.2863 & 0.1380 \\\hline
\textbf{Image} & 7.2821 & 0.1258 & 15.2863 & 0.1380 \\\hline
\textbf{LiDAR} & 5.8630 & 0.1101 & 17.3852 & 0.1830 \\\hline
\textbf{LiDAR + Image} & 8.0177 & 0.1419 & 17.8200 & 0.2041 \\\hline
\textbf{Coordinates + LiDAR} & 6.2575 & 0.1179 & 15.9807 & 0.1867 \\\hline
\textbf{Coordinates + Image} & 7.2821 & 0.1258 & 15.6233 & 0.1567 \\\hline
\textbf{Coordinates + LiDAR + Image} & 6.1026 & 0.1149 & 17.3852 & 0.1830 \\\hline
\hline
\end{tabular}%
\caption{Test error on \dtest dataset across modalities: MAD and MAE in cosines are defined as in \secref{sec:nn-metrics}. Due to a huge network size, the cosine-based encoding performs significantly worse compared to angle-based encoding.}
\label{tab:training-results}
\end{table*}

\subsection{Impact of sensing modalities}\label{sec:results:stageI}
In this section, we analyze the impact of various sensing modalities on the end-to-end performance of our framework.  

We consider all possible combination of sensor information. In case of unimodal setting, we use the individual sensor feature extractors as described in \figref{fig:nn-model-detailed} to predict the beamspace of the channel. In case of multimodal setting, we use the multimodal fusion network to predict the beamspace of the channel. For each case, we train the model using the dataset \dtrain using angle-based encoding and cosine-based encoding of beamspaces. We then evaluate the performance of the trained models on the test dataset \dtest. The test results, in terms of MAD and MAE in cosines, are reported in \tblref{tab:training-results}. 

We observe that the cosine-based encoding performs significantly worse compared to angle-based encoding. This is due to the huge network size to handle 64000-dimensional output when using cosine-based encoding as opposed to 7200-dimensional encoding when using angle-based encoding. 

\begin{figure}[bt]
    \centering
    \subfloat[$U=2$\label{fig:rate-power-modalities-tuned-u2}]{\includegraphics[width=0.49\columnwidth]{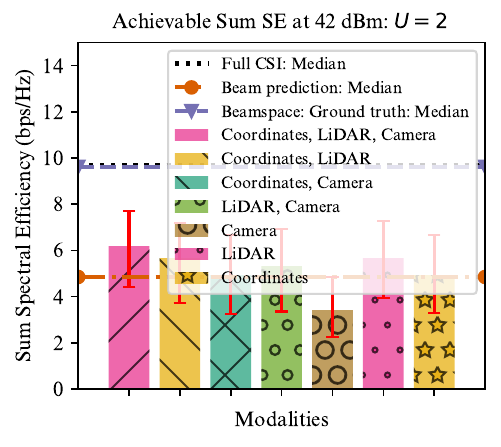}}
    \subfloat[$U=4$\label{fig:rate-power-modalities-tuned-u4}]{\includegraphics[width=0.49\columnwidth]{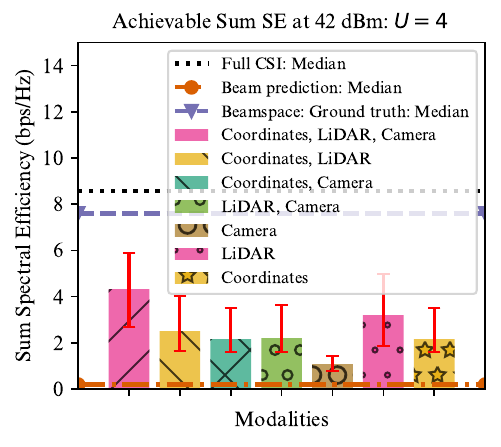}}
    \caption{Comparison of \MU \SE using different combinations of sensor modalities at 42 dBm \EIRP: The error bars show 25\%-tile to 75\%-tile range. The combination of coordinates, LiDAR and camera performs better than other combinations. Notably only using camera images at the BS leads to lower performance compared to the ground-truth baseline which suggest the sensors at the \UEs play significant role in predicting the information about the channel.}
    \label{fig:result-rate-power-modalities}
\end{figure}

We further use these trained models to estimate the sum-\SE of the system. \figref{fig:result-rate-power-modalities} shows the \MU sum-\SE as a function of \EIRP. Compared to the baseline using ground truth-based beamspace representation, we observe that the combination of coordinates, LiDAR and camera performs better than other combinations. Notably only using camera images at the BS leads to lower performance compared to the ground-truth baseline which suggest the sensors at the \UEs play significant role in predicting the information about the channel.

\subsection{Impact of SSCL loss}
\begin{figure}
    \centering
    \subfloat[$U=2$\label{fig:rate-power-contrastive-u2}]{\includegraphics[width=0.49\columnwidth]{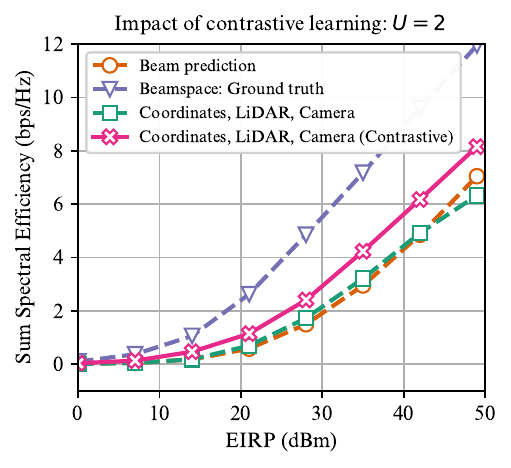}}
    \subfloat[$U=4$\label{fig:rate-power-contrastive-u4}]{\includegraphics[width=0.49\columnwidth]{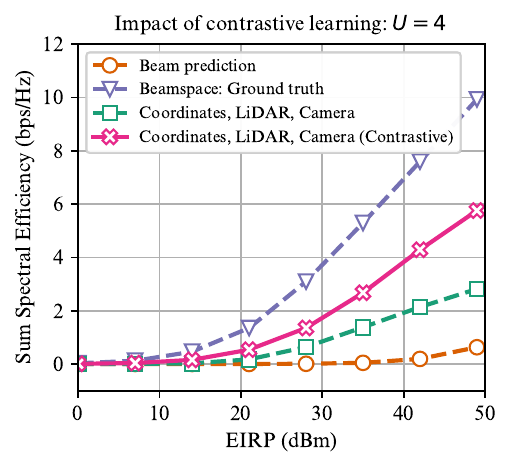}}
    \caption{Defining the impact of \SSCL loss: By contrasting samples based on their similarity score can help recover the relevant features from the sensors for the beamspace prediction, which, in turn, helps with improving achievable \SE using beamspace-aided \MU beamforming strategy.}
    \label{fig:result-rate-contrastive}
\end{figure}

In \figref{fig:result-rate-contrastive}, we show the impact of \SSCL loss on the sum-\SE of \MU communication with $U=2$ and $4$. We observe that the \SSCL loss is a key reason for the success of the proposed multimodal fusion network in predicting the beamspace representation. Since, \SSCL incentivizes feature similarities and dis-similarities during the training of the fusion network, we can recover key features relevant for beamspace estimation from the sensors, and therefore, achieve higher sum-\SE. 

\subsection{Overhead analysis}\label{sec:results:overhead}
To further highlight the benefit of our two-stage approach (sensor-aided beamspace estimation followed by beamspace-aided beamforming) in contrast with the one-stage sensor-aided beam prediction strategy, we discuss the overhead of both strategies. 

Our method only uses sensor data to determine \RF precoder for all users. Therefore, the first phase of \algref{alg:main} incurs no additional overhead on the communication system. 
The 5G NR specification for \mmWave bands allow transmitting at most 64 \SSBs using different \RF precoders. In contrast, our method only uses one \SSB transmission for initial synchronization of \UEs. Considering the 5G NR specifies $5$ ms for the \SSB burst, the \BS, employing sensor-aided \MU beamforming, only consumes $5/64$ ms to transmit one \SSB block with selected \RF precoders. Therefore, the proposed two-stage approach reduces the overhead by the factor of upto $64$. 
Furthermore, the state-of-the-art sensor-aided beam prediction strategy requires roughly $4\ \textrm{ms}$ \textit{per \UE} to determine the \RF precoder with 90\% accuracy~\cite{SalReuRoy:Deep-Learning-on-Multimodal:22}. In comparison, our two-stage approach reduces the overhead by upto 50th factor. 
We again emphasize here that the sum-\SE results presented in this work do not consider the beam selection overhead as we only highlight the benefit in terms of the maximum achievable sum-\SE. In practice, we expect the gain of the proposed two-stage approach for \MU \mmWave beamforming to be even higher if the beam selection overhead are considered in the evaluations. 

While our strategy involves measuring \CSI at the \UEs to design the digital precoder $\Fbb$, this overhead is minimal and unavoidable. It is minimal because for $\Nrf$ RF chains, the \BS needs to transmit only $\Nrf$ \OFDM symbols. Considering 120 kHz sub-carrier spacing for \mmWave communication, each \OFDM symbol has a duration of 8.3 $\mu$s. For example, transmitting to 10 \UEs using 10 \RF chains would only require 0.083 ms, and this time only reduces with increased sub-carrier spacing. This overhead is also unavoidable because the \BS always needs to adjust the digital precoder with the phase-coherent estimate of the channel, regardless of the \RF precoder selection procedure.

\section{Conclusion and Future work}\label{sec:conclusion}
In this paper, we addressed the challenge of establishing efficient links in mmWave MU-MIMO systems by leveraging ubiquitous sensor data. Our results demonstrate that sensor data can reveal not just one dominant path, but the directions of numerous significant paths within the channel. This also highlights the potential of out-of-band sensors for extracting rich channel information to aid advance features of communication systems such as MU-MIMO. 

Looking ahead, we identify three avenues for future research. Firstly, we posit that sensors can also predict the gain of channel paths from the sensor data. Achieving this objective, however, necessitates fundamental changes to the current fusion framework. The revised framework should explicitly integrate LOS classification and develop separate magnitude prediction models for both LOS and non-LOS channel paths. Accurately predicting channel path gains holds the potential to significantly enhance mmWave MU-MIMO systems by facilitating the allocation of varying numbers of data streams and adaptable power levels to each user.

Secondly, realizing the real-world implementation of our approach demands evaluation on real-world datasets containing complete channel measurements. Existing real-world multimodal sensor datasets like DeepSense 6G and e-FLASH unfortunately lack complete channel measurements. Hence, experimental campaigns are required to gather multimodal sensor data alongside comprehensive channel measurements. 

Lastly, for broader deployment of this framework, the adoption of generalizable DNN training alongside site-specific model tuning methodologies needs to be considered. Recent advancements in model-agnostic ML and transfer learning paradigms present promising avenues for addressing these challenges. Overcoming these obstacles, in conjunction with our proposed method, sets the stage for the development of more efficient and resilient MU mmWave communication systems.

\bibliographystyle{IEEEtran}
\bibliography{main.bib}

\end{document}

%% file: glossary.tex
\newacronym{mmWave}{mmWave}{millimeter-wave}
\newcommand{\mmWave}{\gls{mmWave}\xspace}

\newacronym{fr}{FR}{frequency range}

\newacronym{GNSS}{GNSS}{Global Navigation Satellite Systems}
\newcommand{\GNSS}{\gls{GNSS}\xspace}

\newacronym{bs}{BS}{base station}
\newcommand{\BS}{\gls{bs}\xspace}
\newacronym{ue}{UE}{user equipment}
\newcommand{\UE}{\gls{ue}\xspace}
\newcommand{\UEs}{\glspl{ue}\xspace}

\newacronym{dl}{DL}{downlink}

\newacronym{mu}{MU}{multi-user}
\newcommand{\MU}{\gls{mu}\xspace}
\newacronym{su}{SU}{single user}
\newcommand{\SU}{\gls{su}\xspace}
\newacronym{ul}{UL}{uplink}

\newacronym{rf}{RF}{radio frequency}
\newcommand{\RF}{\gls{rf}\xspace}

\newacronym{mimo}{MIMO}{multi-input-multi-output}
\newcommand{\MIMO}{\gls{mimo}\xspace}
\newacronym{miso}{MISO}{multi-input-single-output}
\newcommand{\MISO}{\gls{miso}\xspace}
\newacronym{ofdm}{OFDM}{orthogonal frequency division multiplexing}
\newcommand{\OFDM}{\gls{ofdm}\xspace}
\newacronym{csi}{CSI}{channel state information}
\newcommand{\CSI}{\gls{csi}\xspace}
\newacronym{aoa}{AoA}{angle-of-arrival}

\newacronym{aod}{AoD}{angle-of-departure}
\newcommand{\AoD}{AoD\xspace}
\newcommand{\AoDs}{\glsplural{aod}\xspace}
\newacronym{snr}{SNR}{signal-to-noise ratio}

\newacronym{eirp}{EIRP}{equivalent isotropic radiated power}
\newcommand{\EIRP}{\gls{eirp}\xspace}
\newacronym{inr}{INR}{interference-to-noise ratio}

\newacronym{los}{LOS}{line-of-sight}
\newcommand{\LOS}{\gls{los}\xspace}
\newacronym{nlos}{NLOS}{non-line-of-sight}

\newacronym{rzf}{RZF}{regularized zero-forcing}
\newcommand{\RZF}{\gls{rzf}\xspace}

\newacronym{ssb}{SSB}{synchronization signal block}
\newcommand{\SSB}{\gls{ssb}\xspace}
\newcommand{\SSBs}{\glsplural{ssb}\xspace}
\newacronym{rs}{RS}{reference signal}

\newacronym{se}{SE}{spectral efficiency}
\newcommand{\SE}{\gls{se}\xspace}
\newacronym{mse}{MSE}{mean square error}
\newcommand{\MSE}{\gls{mse}\xspace}
\newacronym{mae}{MAE}{Mean absolute error}

\newacronym{mad}{MAD}{Mean angular distance}

\newacronym{bce}{BCE}{Binary cross entropy}

\newacronym{scl}{SCL}{supervised contrastive loss}
\newcommand{\SCL}{\gls{scl}\xspace}
\newacronym{sscl}{SSCL}{supervised soft-contrastive loss}
\newcommand{\SSCL}{\gls{sscl}\xspace}
\newacronym{rmse}{RMSE}{Root \MSE}

\newacronym{nmse}{NMSE}{Normalized \MSE}

\newacronym{dnn}{DNN}{deep neural network}
\newcommand{\DNN}{\gls{dnn}\xspace}
\newacronym{ml}{ML}{machine learning}
\newcommand{\ML}{\gls{ml}\xspace}
\newacronym{fov}{FoV}{field-of-view}
\newcommand{\fov}{\gls{fov}\xspace}

%% file: notations.tex
\newcommand{\secref}[1]{Section \ref{#1}}

\newcommand{\algref}[1]{Algorithm \ref{#1}}
\newcommand{\figref}[1]{Fig. \ref{#1}}
\newcommand{\tblref}[1]{Table \ref{#1}}

\newcommand{\cc}[1]{\mathcal{#1}}
\newcommand{\bb}[1]{\mathbb{#1}} 

\newcommand{\imj}{\mathsf{j}}

\newcommand{\trans}{\mathrm{T}}
\newcommand{\x}{\mathrm{x}}
\newcommand{\y}{\mathrm{y}}
\newcommand{\vectorize}[1]{\mathrm{vec}(#1)}
\newcommand{\conj}[1]{\bar{#1}}

\newcommand{\quant}[1]{{#1}^{\mathrm{Q}}}
\newcommand{\angdiff}[2]{\mathrm{ADiff}(#1,#2)}

\newcommand{\Nrf}{N_{\mathrm{RF}}}

\newcommand{\Nbsx}{N_{\mathrm{BS}}^\x}
\newcommand{\Nbsy}{N_{\mathrm{BS}}^\y}

\newcommand{\Ns}{N_{\mathrm{S}}}
\newcommand{\Ts}{T_{\mathrm{S}}}

\newcommand{\Fbb}{\bm{\mathsf{F}}_\mathrm{BB}}
\newcommand{\fbb}[1][u]{\bm{\mathsf{f}}_{#1}^{\mathrm{BB}}}
\newcommand{\Frf}{\bm{\mathsf{F}}_\mathrm{RF}}
\newcommand{\frf}[1][u]{\bm{\mathsf{f}}_{#1}^{\mathrm{RF}}}

\newcommand{\dtest}{\texttt{s009}\xspace}
\newcommand{\dtrain}{\texttt{s008}\xspace}